# Generative artificial intelligence and hybrid models to accelerate LES in reactive flows: Application to hydrogen/methane combustion

**Authors:** Xiangrui Zou[1,2], Rodrigo Abadia-Heredia[2], Laura Saavedra[2], Alessandro Parente[3], Rui Xue[1]*, Soledad Le Clainche[2]*

[1] State Key Laboratory for Strength and Vibration of Mechanical Structures, School of Aerospace Engineering, Xi'an Jiaotong University, No.28 West Xianning Road, Xi'an, Shaanxi, 710049, China.

[2] ETSI Aeronautica y del Espacio, Universidad Politécnica de Madrid, Plaza Cardenal Cisneros, 3, Madrid 28040, Spain.

[3] Aero-Thermo-Mechanics Department, Université Libre de Bruxelles, Avenue Franklin D. Roosevelt 50, 1050 Brussels, Belgium.

**Corresponding Author**:

Rui Xue[1]*, Email: ruixue@xjtu.edu.cn

Tel/Fax: +86- 029-82668750

State Key Laboratory for Strength and Vibration of Mechanical Structures, School of Aerospace Engineering, Xi'an Jiaotong University, No.28 West Xianning Road, Xi'an, Shaanxi, 710049, China.

Soledad Le Clainche[2]*, Email: soledad.leclainche@upm.es

Tel: +34 910 67 58 53, Fax: +34 910 67 58 53

ETSI Aeronautica y del Espacio, Universidad Politécnica de Madrid, Plaza Cardenal Cisneros, 3, Madrid 28040, Spain.

**Abstract**

With increasing emphasis on carbon neutrality, accurate and efficient combustion prediction has become essential for the design and optimization of new generation combustion systems. This study established a computational framework by combining large eddy simulation (LES) with a generative machine learning approach which integrates modal decomposition and neural network, enabling fast prediction of hydrogen/methane combustion. A canonical jet-in-hot-coflow burner was selected as the benchmark configuration. LES was performed using eddy dissipation concept model in conjunction with a 17-species and 58-step skeletal mechanism. Reasonable agreement between LES results and experimental data was obtained for temperature and species mass fraction along radial and axial directions, confirming the accuracy of the present LES results. Flow characteristics and flame structures were analyzed, providing a reference for choosing parameters in prediction. Proper orthogonal decomposition (POD) was used to extract dominant flow features, and a hybrid autoregressive model, which combines modal decomposition with a deep learning (POD-DL) was constructed to forecast the temporal evolution of the combustion field. Comparison between the predicted results and LES data, including instantaneous contours, radial distributions, histogram and relative root mean square error, demonstrated a reasonable agreement. The main complexity lies in capturing the chaotic and fine-scale structures inherent to turbulent combustion. To the authors' knowledge, this is the first application of such a hybrid generative model to reactive flow prediction, representing an important step toward using data-driven surrogates to accelerate CFD simulations in combustion research. The proposed approach achieves speed-up ratios of 121 and 845 relative to LES for two tested cases. The implementation is publicly available and will be integrated into the upcoming release of the ModelFLOWs-app (https://modelflows.github.io/modelflowsapp/).

**Keywords**: Large eddy simulation; Reduced order modeling; Deep learning; Data-driven method; Hydrogen combustion; Methane combustion

## 1. Introduction

Combustion plays a fundamental role in modern society, supporting a wide range of applications such as heating systems, transportation, industry processes, and aerospace propulsion [1]. Owing to its extensive applications, combustion has been the subject that has been widely investigated. Among the various investigative methods, numerical simulation has become one of key approaches for gaining insights into combustion phenomena, due to advancements in computational fluid dynamics (CFD) [2]. However, accurately capturing the complex behavior of reacting flows often requires highly refined computational grids and corresponding small time steps. For large-scale systems, it dramatically increases the demand for computational resources, especially when detailed chemical mechanisms involving numerous species and reactions are considered [3–5]. To achieve the goal of carbon neutrality, accurate and efficient predictive methods for capturing combustion evolution and dynamics are key to the development and optimization of combustion systems [6,7]. Such methods enable precise modeling of flame behavior, which is critical for enhancing energy conversion efficiency and reducing carbon emissions. Besides, the inherent complexity of realistic combustion systems makes it particularly challenging to achieve accurate and efficient predictions. Consequently, there is an inevitable need to develop modeling strategies that can reduce computational costs while maintaining acceptable prediction accuracy [8,9].

To address this challenge, reduced order models (ROMs) have gained significant attentions as a means to extract essential flow dynamics and reduce data dimensionality [10]. There are generally two types of ROMs, i.e., projection-based ROMs [11] and data-driven ROMs [12]. Projection-based ROMs firstly construct a low-dimensional subspace based on the full order model (FOM) simulation, onto which the FOM equations are projected to form a set of reduced equations. These methods have been successfully used in various fields, e.g., aerodynamics [13], flow control [14] and combustion [15]. Alternatively, data-driven ROMs can yield accurate description of flow without considering physical equations owing to their equation-free properties. Common approaches include modal decomposition approaches [16], deep neural networks [17], or hybrid methods that combine modal decomposition and deep learning [12,18]. Widely used dimensionality reduction methods include proper orthogonal decomposition (POD) [19,20], dynamic mode decomposition (DMD) [21,22] and other robust extensions. The popularity of these models has further grown with the recent advances in deep learning architectures.

Hybrid ROMs, which combine dimensionality reduction and machine learning, have been applied to various applications such as aerodynamic design and unsteady flow prediction. For example, a non-intrusive hybrid ROM integrating POD and long short-term memory (LSTM) neural network was proposed to identify and simulate fluid dynamics systems without requiring access to the source code of the full model [23]. This model was applied to an ocean gyre and flow past a cylinder, reducing CPU costs by three orders of magnitude while maintaining reasonable accuracy, outperforming traditional POD/radial basis function (RBF) methods. Similarly, two hybrid deep learning-based ROMs (POD-RNN and CRAN) were proposed for predicting unsteady fluid flows, leveraging POD or convolutional neural networks (CNNs) for feature extraction and recurrent neural networks (RNNs) for temporal evolution [24]. Both models perform satisfactorily for the flow past a cylinder. Another framework using autoencoders and U-Net architectures enabled nonlinear ROM construction for CFD simulations, reducing computational cost by nearly two orders of magnitude across five CFD datasets, with minimal loss in accuracy [25]. Le Clainche et al. [12,18,26,27] introduced new predictive hybrid ROMs based on modal decomposition, e.g., POD and DMD, and deep learning methods, e.g. RNN and CNN, by which the complex flows including circular cylinder wake, synthetic jet and two-phase flow were predicted. The prediction accuracy and generalization of the proposed models were verified.

Despite recent advancements in hybrid ROMs, most existing studies have been limited to non-reacting or incompressible problems. Whereas combustion systems introduce additional complexity such as heat release, chemical source terms, and strong coupling between flow and species transport, which may limit the generalizability of ROMs developed solely for fluid dynamics. This makes the prediction of turbulent combustion field particularly challenging. This work introduces a generative artificial intelligence (AI) framework that combines modal decomposition and neural networks to predict the temporal evolution of burning process while leveraging underlying physical principles. The prediction is based on high-fidelity LES results, and modal decomposition was employed to significantly decrease the computational cost.

A well characterized combustion system, the jet-in-hot-coflow (JHC) burner [28], is selected for the present study. The burner operates with hydrogen and methane fuels, which are two promising candidates in the transition toward low-carbon energy systems. Previous numerical studies on the JHC burner have primarily employed the Reynolds-Averaged Navier-Stokes (RANS) method. For example, the influences of oxidizer dilution and fuel jet Reynolds number on the reactive flow field were examined in Ref. [29]. It indicates that the addition of hydrogen is favorable for achieving Moderate or Intense Low-oxygen Dilution (MILD) combustion in syngas mixtures. Parente et al. [30–32] conducted a lot of work on this burner. They proposed calibrated constants for the turbulence and combustion models, which are very meaningful for the further investigations, and validated the feasibility of engineering modeling for NO formation in turbulent flames using finite-rate chemistry combustion models that incorporate detailed mechanisms at acceptable computational costs. Despite RANS-based models own computational efficiency and ability to provide averaged flow fields, these models are limited in capturing the non-equilibrium and transient characteristics of turbulent combustion.

To address this limitation, large eddy simulation (LES) is employed in the present study to generate high-fidelity datasets of the JHC burner. By resolving unsteady flow structures and capturing large-scale turbulent motions, LES provides a more accurate representation of turbulence-chemistry interactions and transient combustion features, making it a more reliable approach for producing training data for ROM development.

The objective of this study is to investigate the combustion properties of JHC burner using LES combined with a generative hybrid AI model. This methodology serves as a proof of concept, demonstrating the potential of generative AI to accelerate CFD simulations in industrial and complex reacting flow applications. The remainder of this article is organized as follows. Section 2 introduces the governing equations used in the LES and combustion modeling, as well as methodology of the POD-DL model. Section 3 presents validation of the LES results to ensure the fidelity of the dataset. Section 4 analyzes the flame structure in the JHC burner and assesses the predictive capabilities of the POD-DL model. Finally, Section 5 concludes the study with a summary of the main conclusions.

## 2. Methodology

This section outlines the methodology adopted in this work, including the governing equations for LES and the combustion model, as well as the theoretical framework of the generative hybrid ROM.

### 2.1. Governing equations for combustion simulation

#### 2.1.1. LES governing equations

The three-dimensional (3D) compressible Navier-Stokes equations are adopted to simulate the reacting flows. The Favre-filtered continuity, momentum, energy and species transport equations are expressed as follows [33]

$$\partial_t(\bar{\rho}) + \nabla\cdot\left(\bar{\rho}\tilde{\mathbf{V}}\right) = 0 \quad (1)$$

$$\partial_t\left(\bar{\rho}\tilde{\mathbf{V}}\right) + \nabla\cdot\left(\bar{\rho}\tilde{\mathbf{V}}\tilde{\mathbf{V}}\right) = -\nabla\bar{P} + \nabla\cdot\left(\bar{\mathbf{S}} - \mathbf{B}\right) \quad (2)$$

$$\partial_t\left(\bar{\rho}\tilde{h}_s\right) + \nabla\cdot\left(\bar{\rho}\tilde{\mathbf{V}}\tilde{h}_s\right) = S\cdot\nabla\tilde{\mathbf{V}} + \partial_t\left(\bar{P}\right) + \nabla P\cdot\tilde{\mathbf{V}} + \nabla\cdot\left(\bar{\mathbf{h}} - \mathbf{b_h}\right) - \sum_{i=1}^{N}\left(\bar{\dot{w}}_i h_{f,i}^{\theta}\right) \quad (3)$$

$$\partial_t\left(\bar{\rho}\tilde{Y}_i\right) + \nabla\cdot\left(\bar{\rho}\tilde{\mathbf{V}}\tilde{Y}_i\right) = \nabla\cdot\left(\bar{\mathbf{j}}_\mathbf{i} - \mathbf{b_i}\right) + \bar{\dot{w}}_i \quad (4)$$

where $\rho$, $\mathbf{V}$, $P$, $h_s$, $Y$ represent the density, velocity, pressure, sensible enthalpy and species mass fraction, respectively. The overbar $\overline{(\bullet)}$ and the tilde $\widetilde{(\bullet)}$ respectively denote unweighted and density-weighted filtered quantities, and $w_i$ and $h_{f,i}^{\theta}$ are respectively the chemical reaction rate and formation enthalpy of $i$th species. And $\bar{\mathbf{S}}$ denotes the filtered viscous stress tensor, which can be expressed as

$$\bar{\mathbf{S}} = 2\mu\tilde{\mathbf{D}}_{dev} \quad (5)$$

where $\mu$ is the dynamic viscosity. $\tilde{\mathbf{D}}_{dev} \equiv \left(\nabla\tilde{\mathbf{V}} + \nabla\tilde{\mathbf{V}}^T\right)/2 - \left(\nabla\cdot\tilde{\mathbf{V}}\right)\mathbf{I}/3$ and $\mathbf{I}$ are deviatoric part of strain rate tensor and identity tensor, respectively. The filtered heat flux $\bar{\mathbf{h}}$ is given as

$$\bar{\mathbf{h}} = \lambda\nabla\tilde{T} \quad (6)$$

where $\lambda$ is the thermal conductivity, and $T$ is the temperature. In Eq. (4), $\bar{\mathbf{j}}_\mathbf{i}$ is expressed as

$$\bar{\mathbf{j}}_\mathbf{i} = D_i\nabla\tilde{Y}_i \quad (7)$$

where $D_i$ indicates mass diffusivity of species $i$. The filtered reaction rate of species $i$ is given as

$$\overline{\dot{w}}_{\mathrm{i}} = MW_{\mathrm{i}} \sum_{j=1}^{M} P_{\mathrm{ij}} \overline{\dot{w}}_{\mathrm{j}} \tag{8}$$

where $MW_{\mathrm{i}}$ is the molecular weight of species *i*. $P_{\mathrm{ij}}$ is the stoichiometric matrix. $\overline{\dot{w}}_{\mathrm{j}}$ represents the reaction rate of the *j*th reaction step.

The gas state equation was utilized to obtain filtered pressure

$$\overline{P} = \overline{\rho} R \tilde{T} \tag{9}$$

where $R$ is the gas constant of mixed species, which can be calculated as

$$R = R_{\mathrm{u}} \sum_{i=1}^{N} Y_{\mathrm{i}} / MW_{\mathrm{i}} \tag{10}$$

where $R_{\mathrm{u}}$ is universal gas constant.

To closure the Eqs. (1)-(4), one-equation sub-grid scale model [34] is employed. The sub-grid scale stress tensor $\mathbf{B}$, flux vectors $\mathbf{b_h}$ and $\mathbf{b_i}$ are calculated as

$$\mathbf{B} = \frac{2}{3} \overline{\rho} k \mathbf{I} - 2\mu_{\mathrm{k}} \tilde{D}_{\mathrm{dev}} \tag{11}$$

$$\mathbf{b_h} = -\frac{\mu_{\mathrm{k}}}{Pr_{\mathrm{t}}} \nabla \tilde{h}_{\mathrm{s}} \tag{12}$$

$$\mathbf{b_i} = -\frac{\mu_{\mathrm{k}}}{Sc_{\mathrm{t}}} \nabla \tilde{Y}_{\mathrm{i}} \tag{13}$$

where $Pr_{\mathrm{t}}$ and $Sc_{\mathrm{t}}$ are respectively the turbulent Prandtl and Schmidt numbers. $\mu_{\mathrm{k}}$ is the sub-grid scale viscosity, which can be calculated as

$$\mu_{\mathrm{k}} = C_{\mathrm{k}} \sqrt{k} \Delta \tag{14}$$

where $C_{\mathrm{k}} = 0.094$ is a constant, $\Delta$ is the grid size, and $k$ is the sub-grid scale kinetic energy, modeled as [34]

$$\partial_{\mathrm{t}} \left( \overline{\rho} k \right) + \nabla \cdot \left( \overline{\rho} \tilde{\mathbf{V}} k \right) = -\mathbf{B} \cdot \tilde{\mathbf{D}} + \nabla \cdot \left( \mu_{\mathrm{k}} \nabla k \right) - c_{\mathrm{E}} \overline{\rho} k^{3/2} \Delta^{-1} \tag{15}$$

where $c_{\mathrm{E}} = 1$ is a dimensionless constant.

### *2.1.2. Combustion model*

The eddy dissipation concept (EDC) model [35] is employed to compute filtered reaction rates. In this model, it assumes that the fluid is split into fine structures where mixing and chemical reactions occur, and surroundings which are governed by large-scale turbulent flow structures. The filtered reaction rate is expressed as

$$\overline{w}_{\mathrm{i}} = MW_{\mathrm{i}} \sum_{j=1}^{M} P_{\mathrm{ij}} \left[ \gamma^{*} \dot{w}_{\mathrm{ij}} \left( \overline{\rho}, Y_{\mathrm{i}}^{*}, T^{*} \right) + \left( 1 - \gamma^{*} \right) \dot{w}_{\mathrm{ij}} \left( \overline{\rho}, Y_{\mathrm{i}}^{0}, T^{0} \right) \right] \tag{16}$$

where $\gamma^*$ is the intermittency factor. And the superscripts * and 0 represent fine structure and surrounding states, respectively.

The KEE kinetic mechanism [36] was used to model the chemical reactions, which includes 17 species and 58 chemical reactions. The detailed parameters of reaction mechanism are listed in Table A1 in Appendix A.

*2.1.3. Numerical method*

The 3D computational domain is subdivided into discrete non-overlapping structured grid cells. The finite volume method (FVM), implemented in the open-source software OpenFOAM-10 [37], is employed to solve the aforementioned equations. Second-order schemes are utilized to discretize the equations. The time derivative term is discretized using the backward scheme. The convective and diffusive terms are discretized using Gauss linear and Gauss linear corrected schemes, respectively, to maintain accuracy on non-orthogonal meshes. Pressure-velocity coupling is handled using the PIMPLE algorithm [38], which combines the features of SIMPLE and PISO methods.

*2.2. POD-DL model*

*2.2.1. Dataset organization*

In the current study, a hybrid generative model combining proper orthogonal decomposition (POD) and deep learning (DL), i.e., POD-DL, is used. Firstly, the data structure organized in the present model is introduced. The high-fidelity LES data are used to construct the training and testing dataset. To include all the variables calculated from LES, the database is organized into tensor form. For 3D condition, the database is a fifth-order tensor **M** with dimensions of $J_1 \times J_2 \times J_3 \times J_4 \times K$ as defined in Ref. [39], for which the components are given as

$$
\begin{aligned}
&\mathbf{M}_{1 j_2 j_3 j_4 k} = \mathbf{m}_1\left(x_{j_2}, y_{j_3}, z_{j_4}, t_k\right) \\
&\mathbf{M}_{2 j_2 j_3 j_4 k} = \mathbf{m}_2\left(x_{j_2}, y_{j_3}, z_{j_4}, t_k\right) \\
&\cdots \\
&\mathbf{M}_{j_1 j_2 j_3 j_4 k} = \mathbf{m}_{j_1}\left(x_{j_2}, y_{j_3}, z_{j_4}, t_k\right) \\
&\cdots \\
&\mathbf{M}_{J_1 j_2 j_3 j_4 k} = \mathbf{m}_{J_1}\left(x_{j_2}, y_{j_3}, z_{j_4}, t_k\right)
\end{aligned} \tag{17}
$$

where $J_1$ represents the number of components considered, and $J_2$, $J_3$ and $J_4$ are respectively the discrete values in $x$, $y$, and $z$ coordinates. And $k$ denotes the index of snapshot at time instant $t_k$. For operational convenience, tensor data is converted into a matrix form, which can be achieved through reshaping. The dataset in matrix form can be expressed as

$$
\mathbf{M}_1^K = \left[\mathbf{m}_1, \mathbf{m}_2, \ldots, \mathbf{m}_k, \mathbf{m}_{k+1} \ldots, \mathbf{m}_{K-1}, \mathbf{m}_K\right] \tag{18}
$$

where $\mathbf{M}_1^K \in$ J×K, in which $J = J_1 \times J_2 \times J_3 \times J_4$, $\mathbf{m}_k = \mathbf{m}(t_k)$ is the snapshot at time instant $t_k$. And $K$ is the total number of snapshots collected. To raise the prediction efficiency, we predict each variable in parallel with different CPU cores.

### *2.2.2. POD reduction method*

One of the advantages of ROMs is to reduce the data dimensionality, thereby significantly decreasing the dependence on computational resource. The POD approach, introduced by Lumley [40], is widely used to extract coherent structures form turbulent flows by decomposing data into orthogonal modes that maximize energy content. The classical POD approach relies on covariance analysis, but becomes computationally intensive for high-dimensional data. To address this, Sirovich [41] proposed using singular value decomposition (SVD), which efficiently computes POD modes, especially in large-scale problems.

In fluid dynamics, SVD identifies dominant spatial patterns (POD modes) and their temporal evolution through corresponding coefficients. SVD is one of the primary techniques for performing POD. The spatio-temporal data $\mathbf{m}(x, y, z, t)$ can be decomposed into proper orthogonal spatial modes (i.e., POD modes) $\boldsymbol{\phi}_\mathbf{n}(x, y, z)$, weighted by temporal coefficients $\mathbf{c}_\mathbf{n}(t)$, as follows

$$\mathbf{m}(x,y,z,t) \simeq \sum_{n=1}^{N} \mathbf{c_n}(t)\mathbf{\Phi_n}(x,y,z) \tag{19}$$

Based on SVD algorithm, snapshots matrix $\mathbf{M}_1^{\mathrm{K}}$ is decomposed into a set of orthogonal spatial modes **W** (i.e., the SVD or POD modes), temporal modes **T**, and a diagonal matrix of singular values $\mathbf{\Sigma}$, as follows:

$$\mathbf{M}_1^{\mathrm{K}} \simeq \mathbf{W\Sigma T}^{\mathrm{T}} \tag{20}$$

where $()^{\mathrm{T}}$ represents the transpose of matrix. $\mathbf{\Sigma} = \mathrm{diag}(s_1, s_2, s_3, ..., s_{\mathrm{N}})$ consists of singular values $s_{\mathrm{i}}$ that are sorted in descending order, and $\mathbf{W}^{\mathrm{T}}\mathbf{W} = \mathbf{T}^{\mathrm{T}}\mathbf{T}$ is the unit matrix of order $N$, in which $N$ represents the retained number of singular values. The selection of singular values is usually performed by defining a tolerance $\varepsilon_{\mathrm{SVD}}$, which follows the relationship $s_{\mathrm{N+1}}/s_1 \le \varepsilon_{\mathrm{SVD}}$. Singular values whose ratios to the first singular value are below the specified tolerance are ignored, leading to dimensionality reduction. The reconstruction error introduced by this process can be assessed using the relative root mean square error (RRMSE), computed as

$$RRMSE = \frac{\left\| \mathbf{M}_1^{\mathrm{K}} - \mathbf{W\Sigma T}^{\mathrm{T}} \right\|_2}{\left\| \mathbf{M}_1^{\mathrm{K}} \right\|_2} \tag{21}$$

where $\|\bullet\|_2$ is the Frobenius norm.

To estimate the number of retained modes whether they are enough to capture the state of flow field, the decay and cumulative energy of the most energetic modes are analyzed. The cumulative energy ratio is computed as

$$E_\kappa = \frac{\sum_{\mathrm{i}=1}^{\kappa} s_{\mathrm{i}}}{\sum_{\mathrm{i}=1}^{\mathrm{K}} s_{\mathrm{i}}} \tag{22}$$

where $s_{\mathrm{i}}$ is the $i$th singular value of $\mathbf{\Sigma}$. And the retained mode number $\kappa$ is varied to yield the trend of cumulative energy. Additionally, the contours of different variables, e.g., $T$, $Y_{\mathrm{i}}$, are reconstructed by retaining different modes, which are compared with the original LES results, to select the necessary number of retained modes.

*2.2.3. Deep learning prediction model*

In this section, we combine POD with deep learning (DL) (POD-DL) model to predict the future snapshots. A long short-term memory (LSTM) based recurrent neural network (RNN) [12] was adopted to forecast the evolution process of the weighted POD coefficients, namely $\mathbf{\Sigma T}^{\mathrm{T}}$. Centering and scaling were performed to improve the stability and effectiveness of the decomposition and subsequent neural network training. Centering refers to the removal of the temporal mean from each spatial location. Specifically, the mean value across all snapshots is subtracted from each row of $\mathbf{M}_1^{\mathrm{K}}$ before applying POD, emphasizing deviations or perturbations relative to the mean flow and ensuring that the POD modes capture the dominant fluctuating structures rather than the stationary component of the field. Here we define the centered POD coefficients as $\hat{\mathbf{T}}$. To normalize the temporal coefficients obtained from POD, each row of the coefficient matrix was standardized by subtracting its mean and dividing by its standard deviation. Mathematically, this process can be expressed as:

$$\widehat{\mathbf{T}} = \frac{\hat{\mathbf{T}} - \overline{\hat{\mathbf{T}}}}{\boldsymbol{\sigma}} \tag{23}$$

where $\overline{\hat{\mathbf{T}}}$ and $\boldsymbol{\sigma}$ are the temporal mean and standard deviation of matrix $\hat{\mathbf{T}}$, respectively.

The prediction task is formulated as a sequence-to-sequence regression problem, where the goal is to estimate the POD coefficients at time $t$+1, denoted as $\widehat{\mathbf{T}}_{\mathrm{t+1}}$, based on the previous $q$ temporal snapshots $\widehat{\mathbf{T}}_{\mathrm{t-q+1}}, \widehat{\mathbf{T}}_{\mathrm{t-q+2}}, \ldots, \widehat{\mathbf{T}}_{\mathrm{t}}$. Each snapshot corresponds to a column in the matrix $\widehat{\mathbf{T}}$, which encodes the temporal evolution of the reduced-order state.

The neural network architecture is composed of three sequential components. First, a multi-layer LSTM block with two stacked layers and a hidden size of 100 processes the input sequence of POD coefficients, capturing temporal dependencies across snapshots. The final hidden state of the LSTM is then passed through a two-layer fully connected network with ReLU activation function and 200 neurons per layer, which extracts nonlinear temporal features. Lastly, a linear output layer maps these features to predict the modal coefficients at the next time step.

The model is trained using the Adam optimizer with a learning rate of $1\times10^{-3}$, and batch size of 12. The input sequence length is fixed to 10 consecutive snapshots (i.e., $q$ = 10) for all training and prediction tasks. The training objective is to minimize the mean squared error (MSE) between predicted and ground-truth modal coefficients, the MSE is computed for each time prediction, expressed as

$$MSE_{\text{loss}}\left(t\right)=\frac{1}{N}\left\|\widehat{\mathbf{T}}_{\text{t,pred}}-\widehat{\mathbf{T}}_{\text{t,CFD}}\right\|_2 \tag{24}$$

where $N$ represents the number of retained modes. The global loss is calculated by averaging the local loss, given as

$$MSE_{\text{loss}}=\frac{1}{K_\alpha}\sum_{K_\alpha}MSE_{\text{loss}}\left(t\right) \tag{25}$$

where $K_\alpha$ is the number of samples for validation or test datasets.

To evaluate the performance of the POD-DL model, RRMSE between the predicted and LES results is also utilized, expressed as

$$RRMSE(t)=\frac{\left\|\mathbf{M}_{\text{t,CFD}}-\mathbf{M}_{\text{t,Pred}}\right\|_2}{\left\|\mathbf{M}_{\text{t,CFD}}\right\|_2} \tag{26}$$

where $\mathbf{M}_{\text{t,CFD}}$ and $\mathbf{M}_{\text{t,Pred}}$ are the original CFD and predicted results at the same time instant $t$, respectively.

## 3. Computational geometry and model validation

The flame considered in this study is a jet-in-hot-coflow (JHC) burner, proposed by Dally et al. [28]. A schematic of the burner configuration is shown in Fig. 1. The burner is composed of a central fuel jet with an inner diameter of 4.25 mm ($D$), and an annulus hot coflow inlet with an inner diameter of 82 mm (i.e., 19.3$D$). During the experiments, the entire burner assembly is mounted in a wind tunnel. Therefore, room temperature sounding air is provided around the hot coflow. The central fuel jet supplies a mixture composed of hydrogen and methane at 305 K, with a Reynolds number of 9482. In this study, the hydrogen content in the fuel mixture is 11.1% by mass. Regarding the hot coflow, the mixture of $O_2$, $N_2$, $H_2O$ and $CO_2$ is provided with mass fractions of 9%, 79%, 6.5% and 5.5%, respectively. The temperature and velocity of hot coflow inlet are 1300 K and 3.2 m/s, respectively. The temperature and velocity at air inlet are 292 K and 3.2 m/s, respectively.

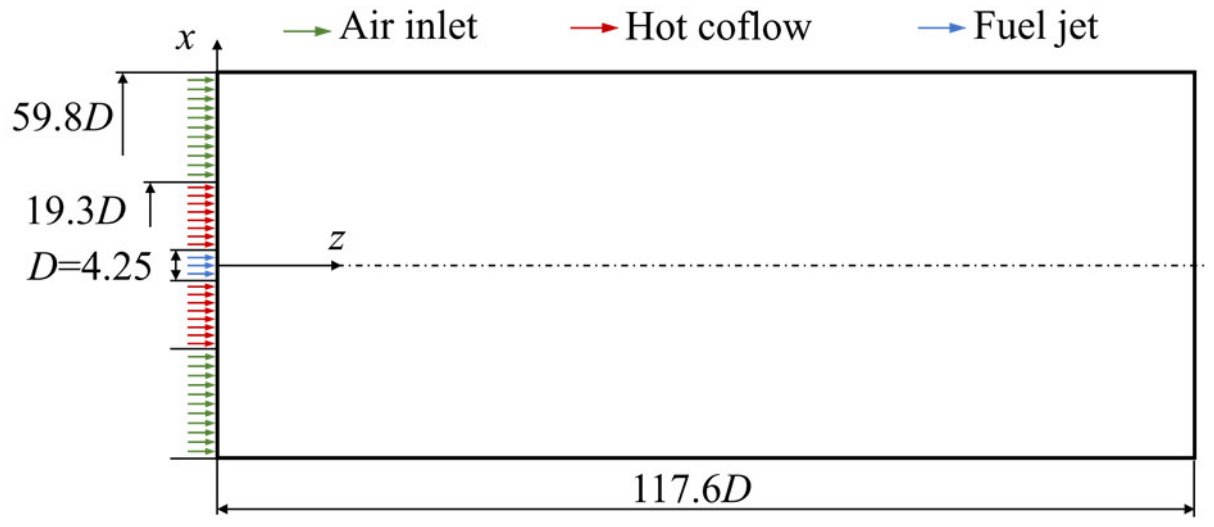

**Fig. 1.** Configuration and dimensions of computational domain.

The combustion process of the JHC burner was simulated using LES. Figure 2 shows the computational mesh, which consists of hexahedral cells with local refinement near the central fuel jet to capture detailed flow structures. The total number of the mesh cells is 1.96 million, which is of the same order of magnitude as those in Refs. [42,43]. Boundary conditions of fuel, hot coflow and air inlets are set as fixed values for velocity, temperature and mass fraction of species. A wave-transmissive boundary was employed at the outlet for pressure, which is a non-reflective condition allowing wave disturbances to pass freely. Combustion was initiated by imposing a high-temperature region (2000 K) just downstream of the fuel inlet.

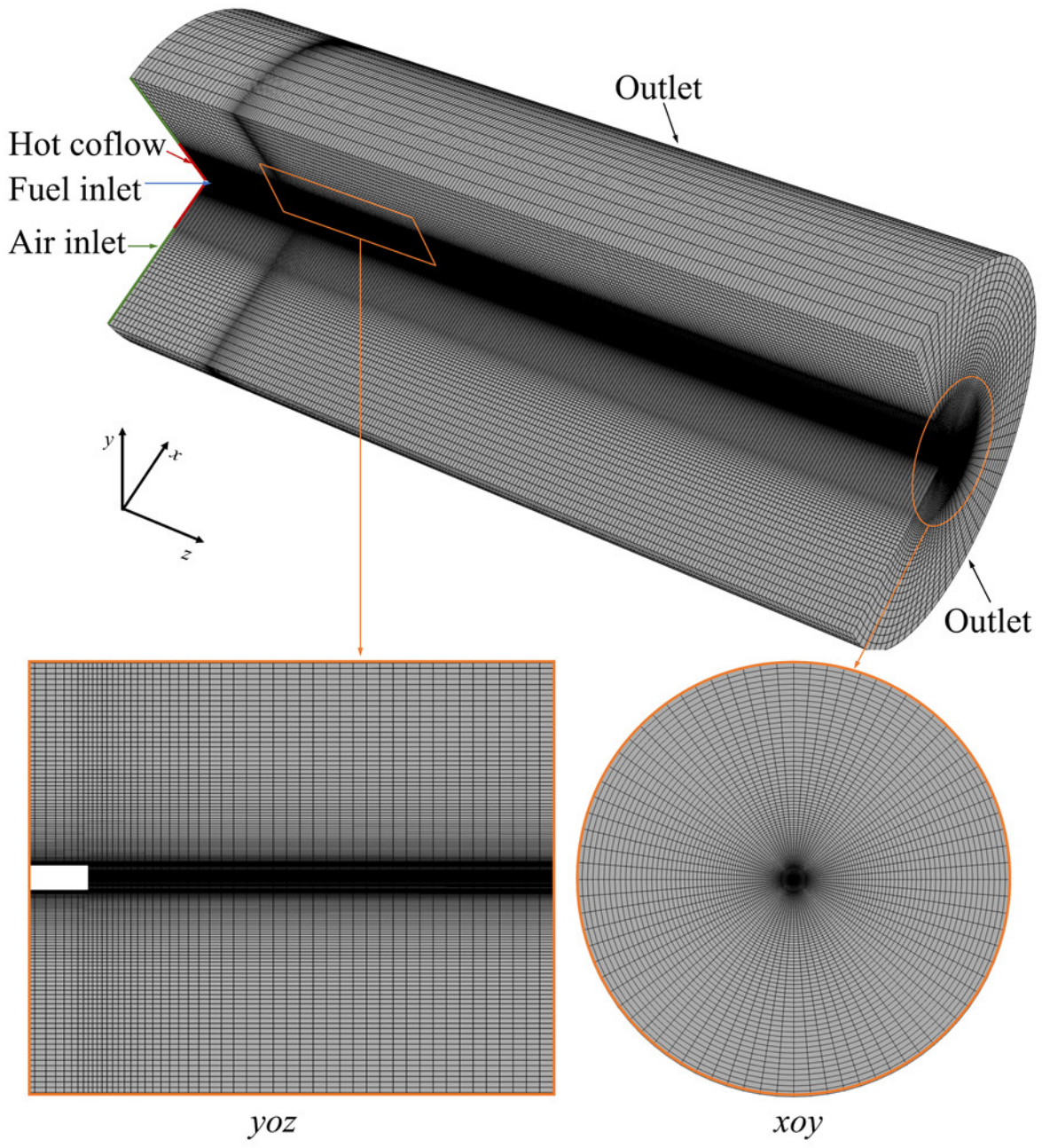

**Fig. 2.** Computational mesh and applied boundary conditions.

To validate the accuracy of the LES results, simulated results were compared against experimental data for the HM3 case from Dally et al. [28]. The validation was performed by examining temperature and species mass fractions (CO, OH and $H_2O$) along the radial direction at axial positions *Z*/*D* = 7.1 and *Z*/*D* = 28.2. Temporal averaging was applied during the simulation, and spatial average over the XOY plane was used for comparison, as the fields are spatially non-uniform. The comparison between LES results and experiment data is shown in Fig. 3. As can be seen, the temperature peaks and overall variation trend are properly captured. Species distributions reasonably agree with the measurements, although some discrepancies in peak values exist. The mass fraction of CO is slightly overpredicted, while OH is slightly underpredicted. Nevertheless, the results indicate satisfactory agreement between LES and experimental data for both temperature and major species.

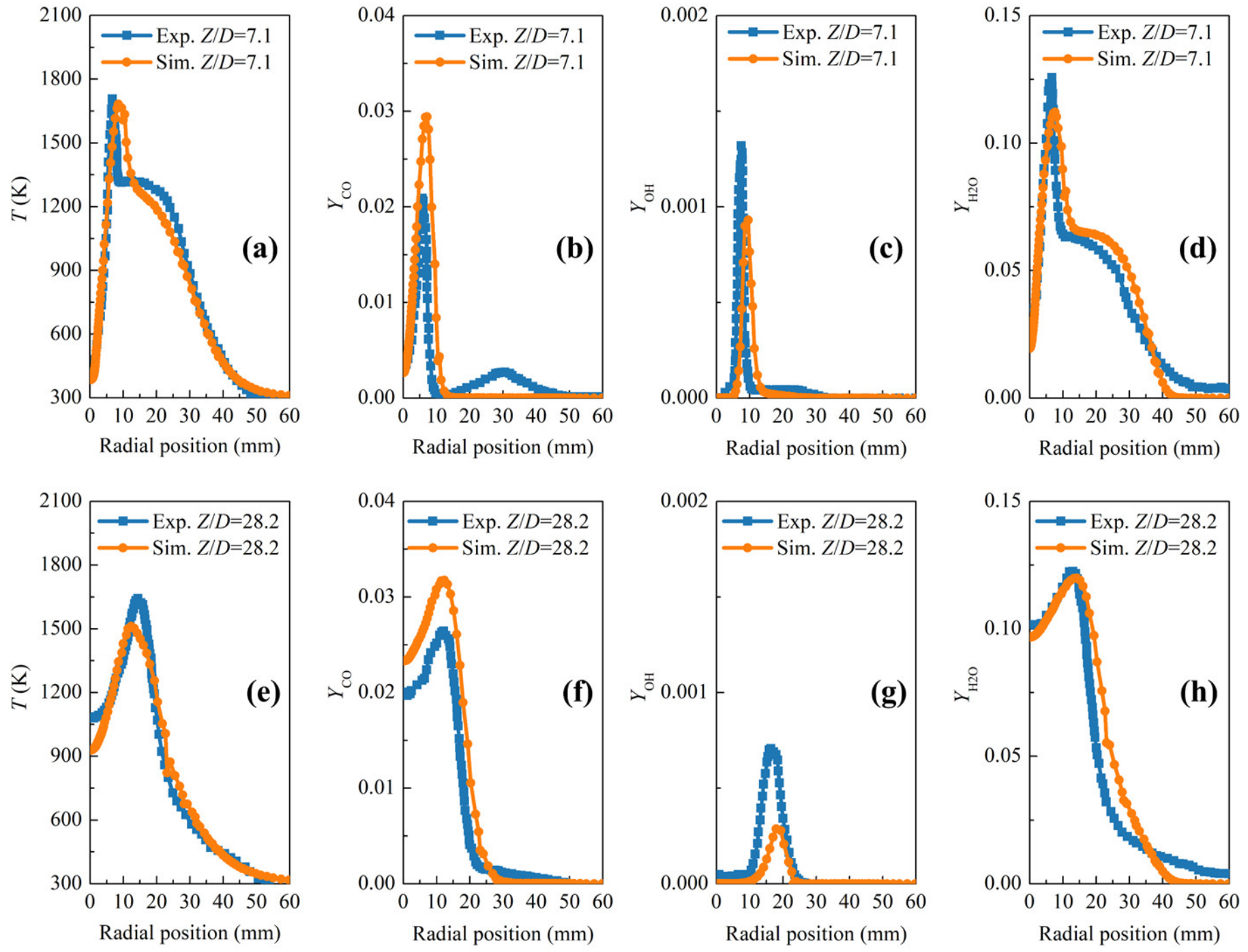

**Fig. 3.** Comparisons of temperature, mass fractions of CO, OH and $H_2O$ along radial direction at *Z*/*D* = 7.1 and *Z*/*D* = 28.2.

Further validation was conducted by comparing the statistical mean and root mean square (RMS) of the temperature along the jet centerline with experimental results given in Ref. [43]. As shown in Fig. 4, the mean temperature follows the experimental trend up to *Z*/*D* = 20. Beyond this location, the simulation underpredicts the temperature. The simulated RMS temperature is approximately 100 K, compared to an experimental value of around 150 K, indicating an underestimation of temperature fluctuations. This can be attributed to the given smaller RMS fluctuation scale of turbulent velocity inlet. Despite these deviations, the results demonstrate that the current LES simulation achieves acceptable accuracy in predicting both mean and fluctuating thermal fields.

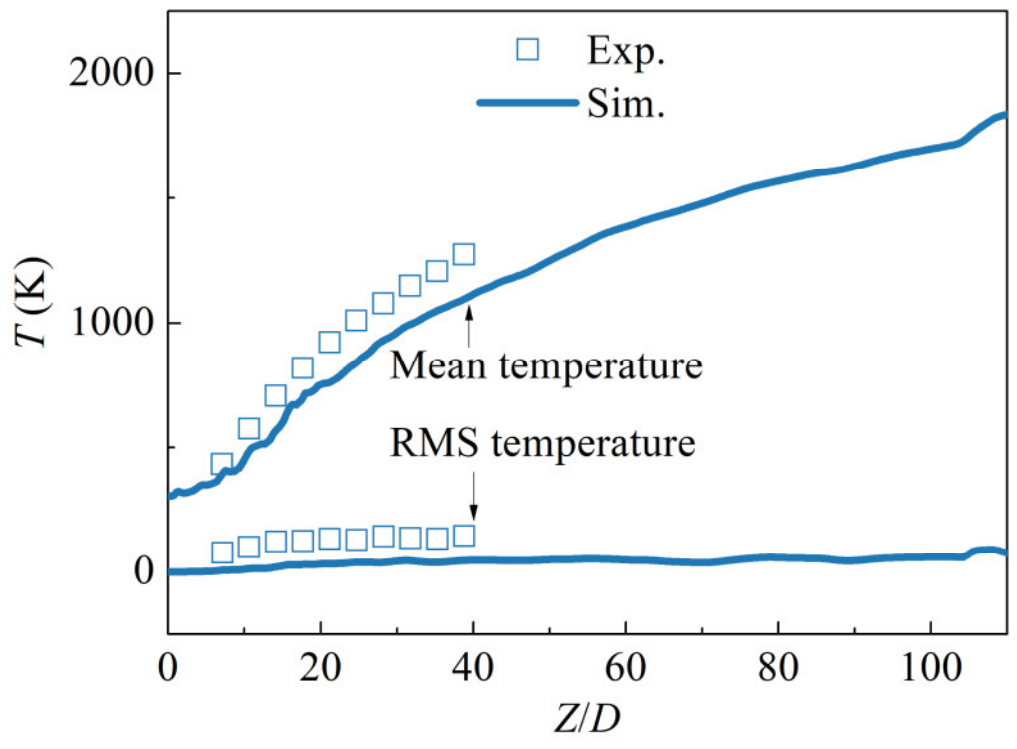

**Fig. 4.** Comparison of statistical and measured mean and RMS temperature profiles along the jet centerline.

## 4. Results and discussion

### *4.1. Flow and flame structures*

To investigate the flow and flame structures, the distributions of velocity, temperature and species mass fractions are analyzed based on LES results. Figure 5 (a) and (b) exhibit the instantaneous mass fractions of $H_2$ and $CH_4$ on the XOZ half plane, respectively. The initial mass fractions at inlet are 11% for $H_2$ and 89% for $CH_4$. As can be seen, both hydrogen and methane diffuse outwards from the central jet and react with oxygen in the surrounding hot coflow. At the axial location of $Z/D$ = 70, the mass fractions of $H_2$ and $CH_4$ decrease significantly, falling below 0.01 and 0.1, respectively. The instantaneous and time averaged temperature fields on the XOZ half plane is shown in Fig. 5 (c) and (d), respectively. The flame is slightly lifted at the exit of fuel jet, and exhibits an inverted conical shape. Combustion initiates in the mixing layer where the fuel and oxidizer blend and then propagates radially outward. The distribution of OH radicals, shown in Fig. 5 (e) and (f), indicates a distinct reaction zone where the temperature is elevated compared to the adjacent fuel and hot coflow streams. From the distribution of $H_2O$ as given in Fig. 5 (g) and (h), it is found that a wavy flame front is formed in the instantaneous field. The interaction between the central fuel jet and surrounding coflow generates a mixing region that sustains combustion. In the downstream, the temperature rise in fuel-rich region is mainly due to the entrainment and mixing with hot combustion products. Beyond *Z*/*D* > 70, variation of both temperature and $H_2O$ mass fraction stabilizes, indicating the completion of major combustion processes.

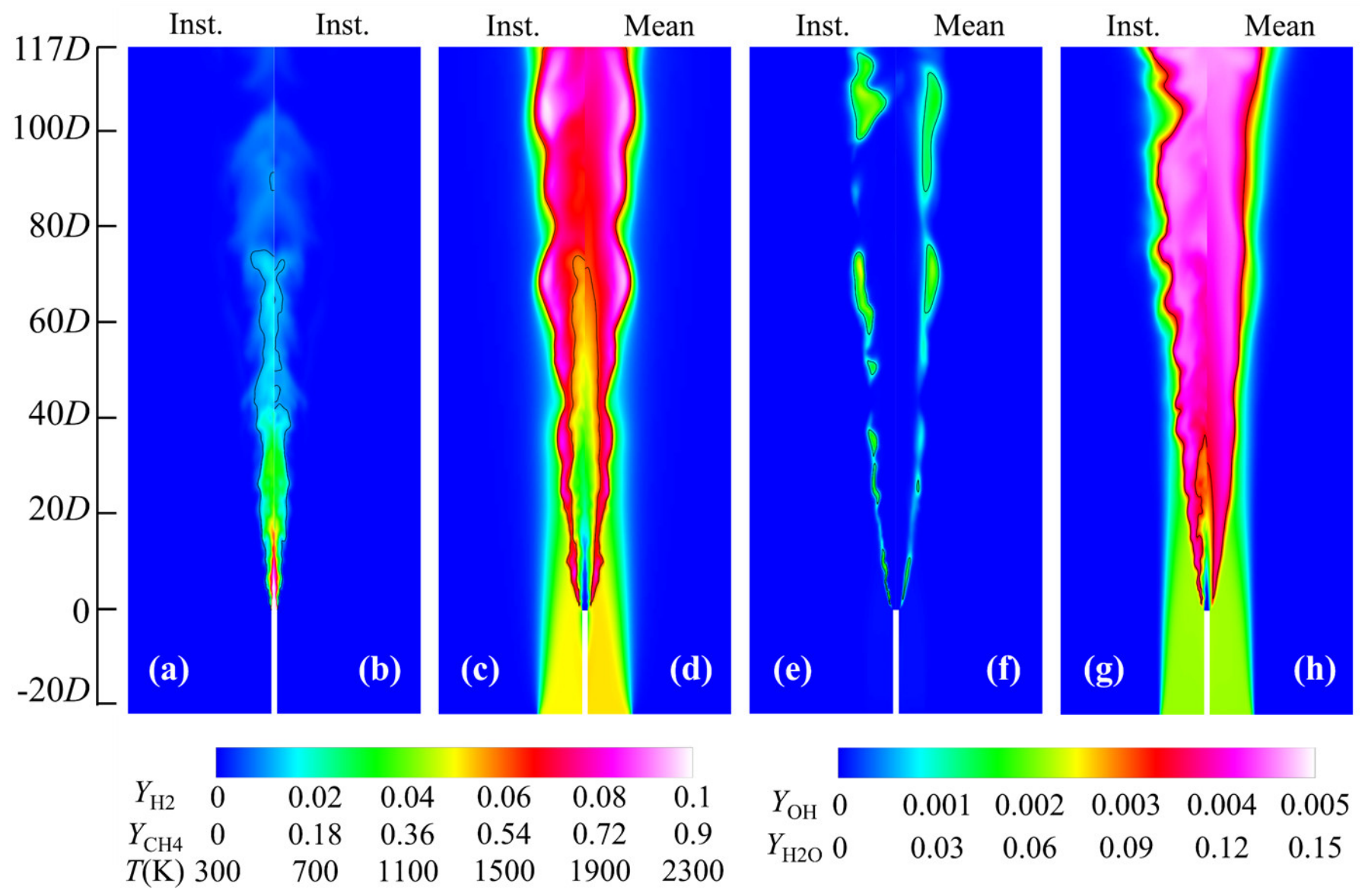

**Fig. 5.** Instantaneous mass fractions of (a) $H_2$ and (b) $CH_4$; (c, d) instantaneous and mean temperature; (e, f) instantaneous and mean OH mass fraction; (g, h) instantaneous and mean $H_2O$ mass fraction. Black isolines represent temperature at 1500 K and mass fractions of $H_2$, $CH_4$, OH, and $H_2O$ at 0.01, 0.1, 0.001, 0.1, respectively.

Figure 6 illustrates the distributions of velocity and vorticity in both axial and radial directions. In Fig. 6 (a), the streamwise velocity distribution is shown, along with a contour line at 3.5 m/s. It can be observed that the velocity of the central jet gradually decreases along the axial direction, while the velocity in the surrounding mixing layer increases due to jet entrainment. Outside this region, the velocity remains around 3.2 m/s. Figure 6 (b) exhibits the radial velocity field, with the 0 m/s contour line representing the transition between different velocity directions. The radial velocity ranges between -4 m/s and 4 m/s. Figure 6 (c) shows the distribution of the axial vorticity. It can be noticed that localized high-vorticity regions near the jet centerline are generated, indicating the presence of strong vortex structures along the axial direction, likely induced by velocity gradients. Moreover, radial vorticity shown in Fig. 6 (d) exhibits high values in corresponding regions, possibly enhanced by thermal gradients near the flame front.

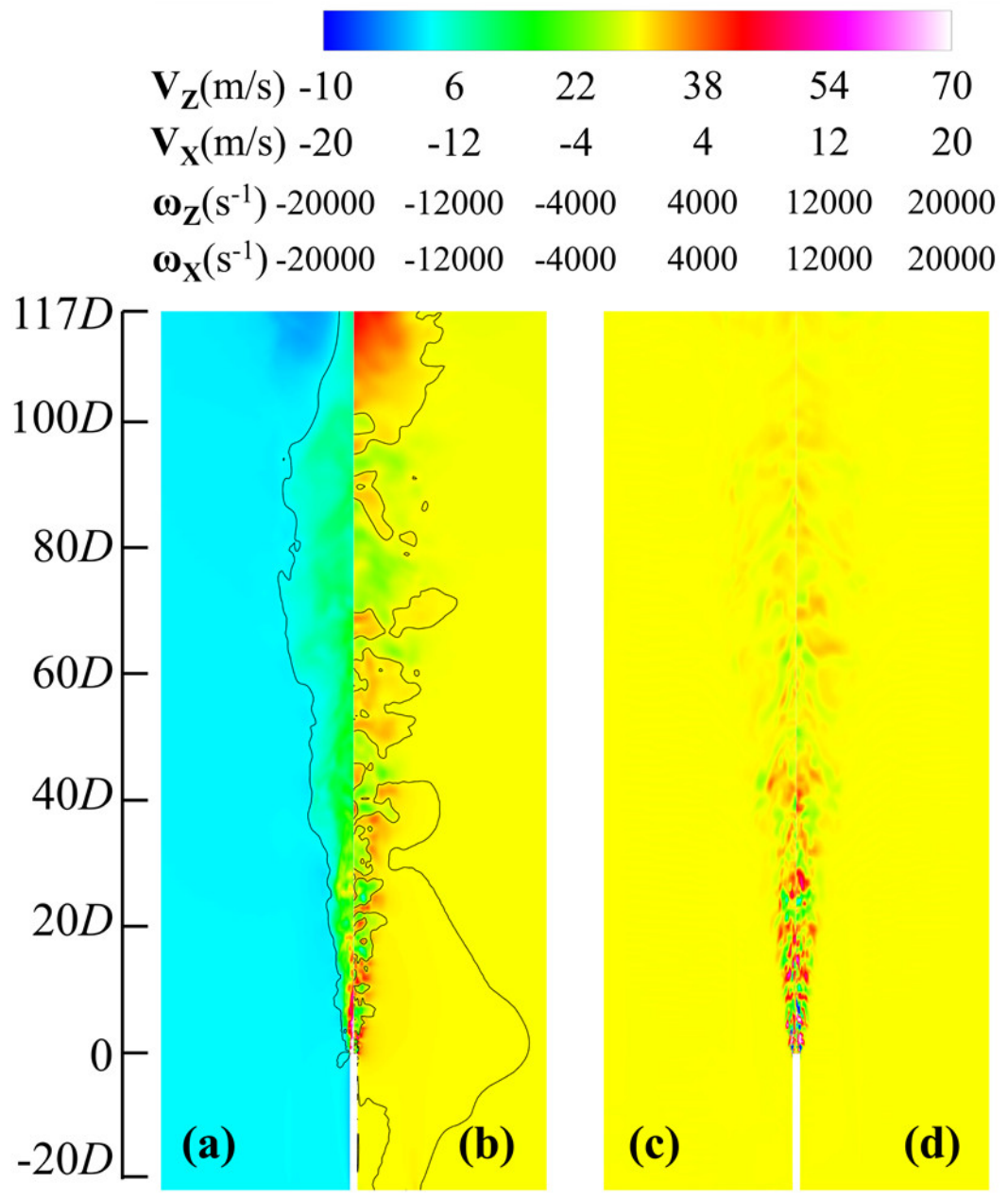

**Fig. 6.** Distributions of (a) streamwise and (b) radial velocity, and (c) streamwise and (d) radial vorticity. Black isolines denote axial and radial velocities at 3.5 m/s and 0 m/s.

To determine the parameters selected to create snapshots matrix for POD and model training, a frequency analysis was conducted using the Fast Fourier Transform (FFT) applied to velocity data monitored at a characteristic position (*D*, 0, 23.5*D*). Figure 7 (a) and (b) present the temporal evolution of streamwise and radial velocity components, along with their FFT results under various sampling intervals. For time intervals of $1\times10^{-5}$ s and $2\times10^{-4}$ s, the velocity profiles exhibit consistent fluctuations, indicating that flow dynamics are adequately resolved. When the time interval increases to $1\times10^{-3}$ s, small scale fluctuations are smoothed out, but the overall flow behavior remains accurately captured. The corresponding FFT results are shown in Fig. 7 (c) and (d). At smaller time intervals ($1\times10^{-5}$ s and $2\times10^{-4}$ s), the dominant frequencies observed are 130 Hz for the streamwise component and approximately 677 Hz for the radial component. These frequencies shift downward when the sampling interval increases to $1\times10^{-3}$ s, due to resolution loss at high frequencies. The spectral truncation for the time interval of $1\times10^{-3}$ s in the FFT plots is attributed to the limited sampling duration (0.2 s). Based on these observations, a time interval of $2\times10^{-4}$ s is deemed sufficient to capture the fine-scale turbulent structures in the flow field, while a time interval of $1\times10^{-3}$ s is adequate for representing the overall flow features.

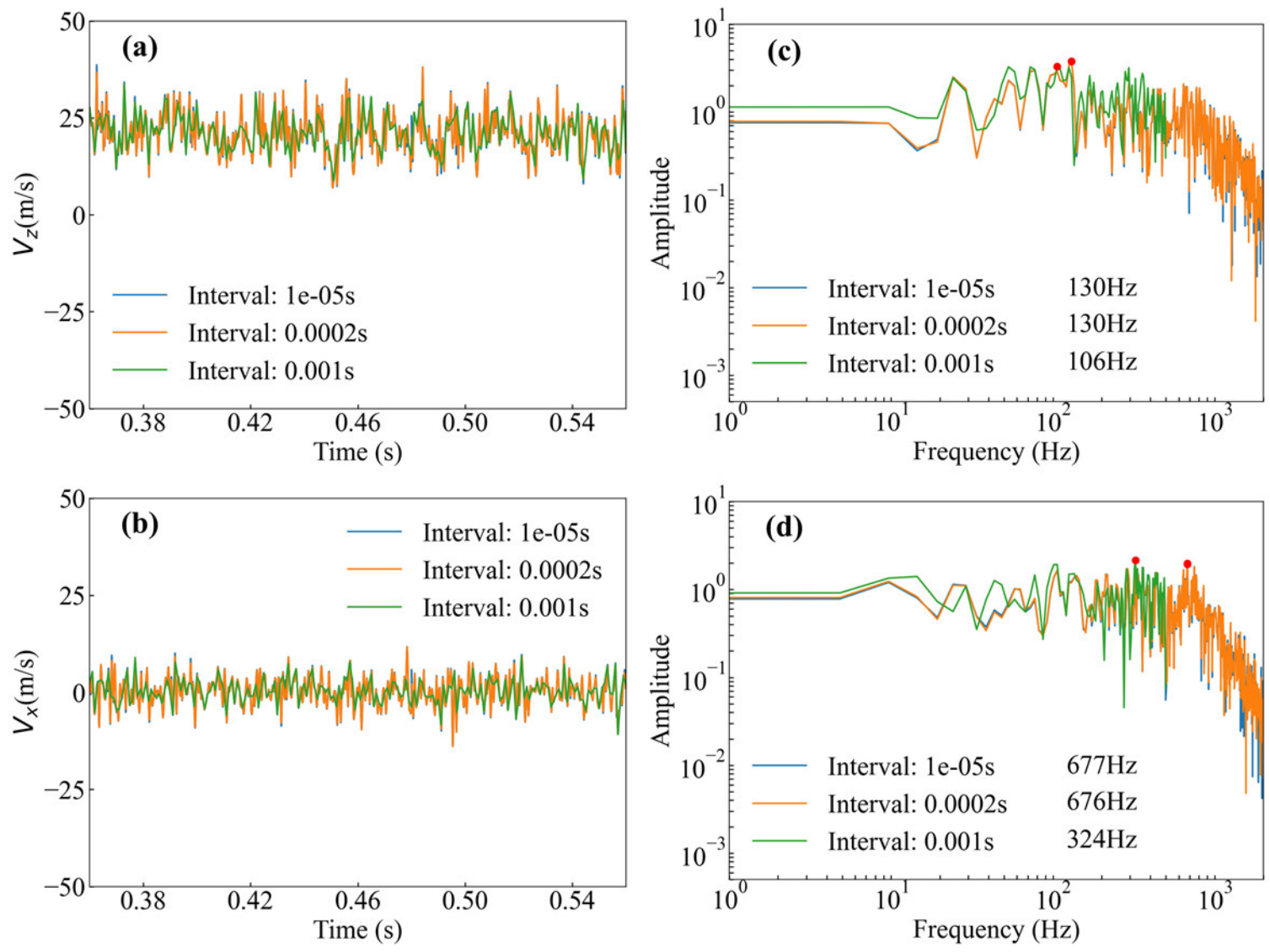

**Fig. 7.** Time histories of (a) streamwise and (b) radial velocity, and FFT spectra of (c) streamwise and (d) radial velocities at a characteristic monitoring point.

*4.2. POD analysis*

After completing the LES calculations, the simulation data were organized into a series of temporal snapshots. In this study, a total of 500 snapshots were collected with the time interval of $2\times10^{-4}$ s, and 200 snapshots were extracted for time interval of $1\times10^{-3}$ s, corresponding to physical time range of 0.4s-0.5s and 0.36s-0.56s, respectively. The former dataset was analyzed in this section since it sustains the same resolution with the original data. The first half of the dataset (250 snapshots) was used for training, while the remaining half was reserved for testing. To determine the appropriate number of modes required for accurately representing the original data, the cumulative energy of singular values and reconstruction accuracy with different numbers of retained modes were examined. Although this analysis is applicable to all simulated variables, a representative analysis was conducted for four key quantities: temperature, mass fractions of OH, CO and $H_2O$. Figure 8 (a) illustrates the decay of the first 100 most energetic modes, while Fig. 8 (b) shows the cumulative energy ratio, as defined by Eq. (22), plotted against the number of retained singular values. It is observed that for OH and CO mass fractions, the first 30 and 50 modes capture approximately 60% and 70% of the total energy, respectively. In comparison, the cumulative energy ratios for temperature and $H_2O$ mass fraction are about 10% lower.

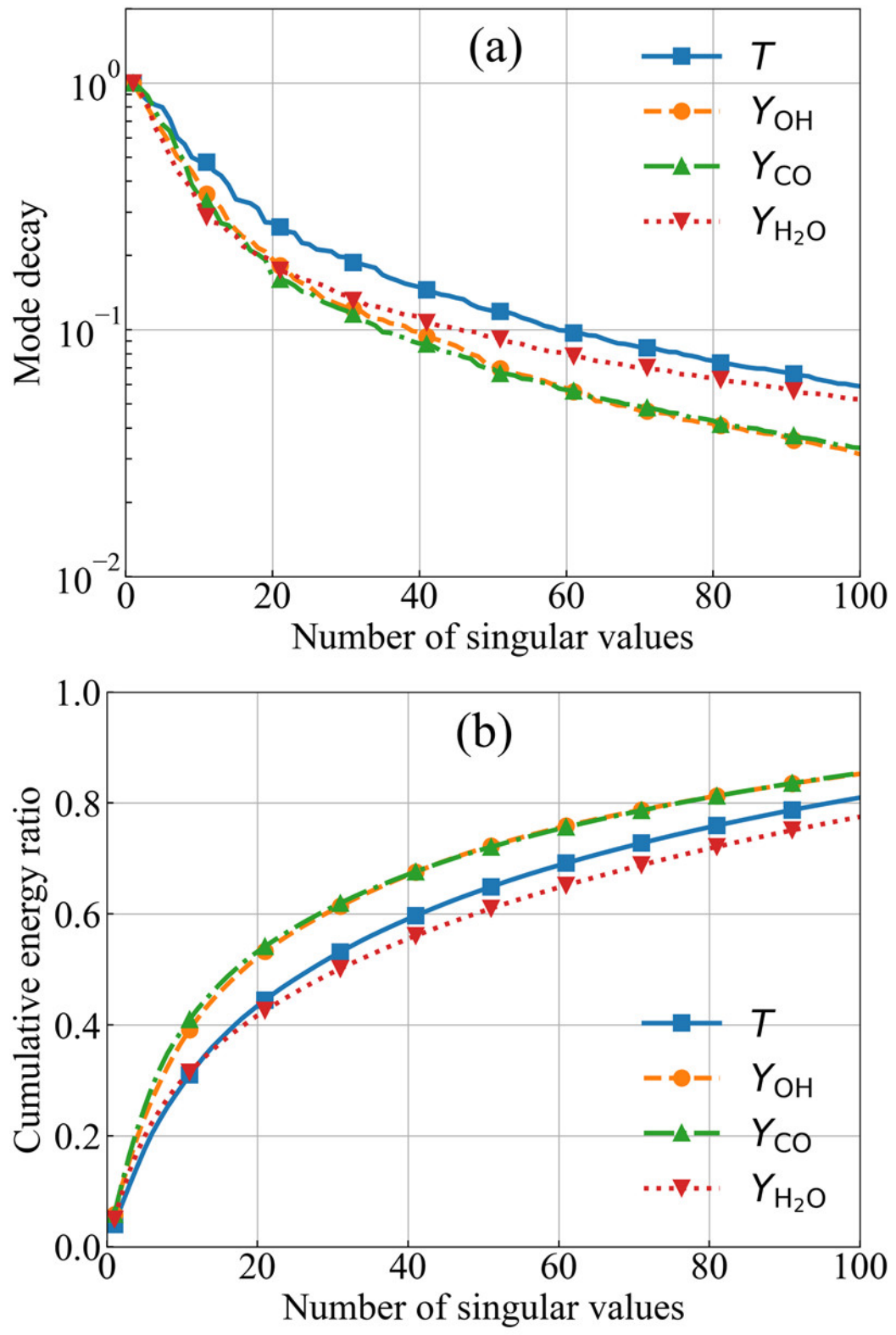

**Fig. 8.** Decay of mode energy and cumulative energy ratio for the first 100 modes of four representative variables: temperature, mass fractions of OH, CO, $H_2O$.

To further assess the representational fidelity of different modal truncation levels, POD was performed on the training dataset. The reconstruction of the temperature field and CO mass fraction was carried out using 10, 30 and 50 retained modes. A comparison between the reconstructed fields and the corresponding ground truth for a representative snapshot is presented in Fig. 9 (a)-(h). And Fig. 9 (j)-(l) present the difference between reconstructed and ground truth for temperature. When only 10 modes are retained, noticeable discrepancies are observed, fine structures are lost, and there exists difference in isoline between reconstruction and ground truth. Increasing the number of modes to 30 significantly improves the accuracy, capturing the major flow features. With 50 modes, the reconstructed fields closely resemble the ground truth, preserving both large-scale and fine-scale structures. The difference shown in Fig. 9 (j-l) confirm these results. Since the input of neural network is the temporal modes of POD, it is essential that the POD reconstruction faithfully represents the original data to guarantee that the model predicts the small-scale dynamics, generally related with the less amplitude (energetic) POD modes. Therefore, retrieving at least 30 modes is considered necessary to ensure reliable training and predictive performance of the neural network in this study.

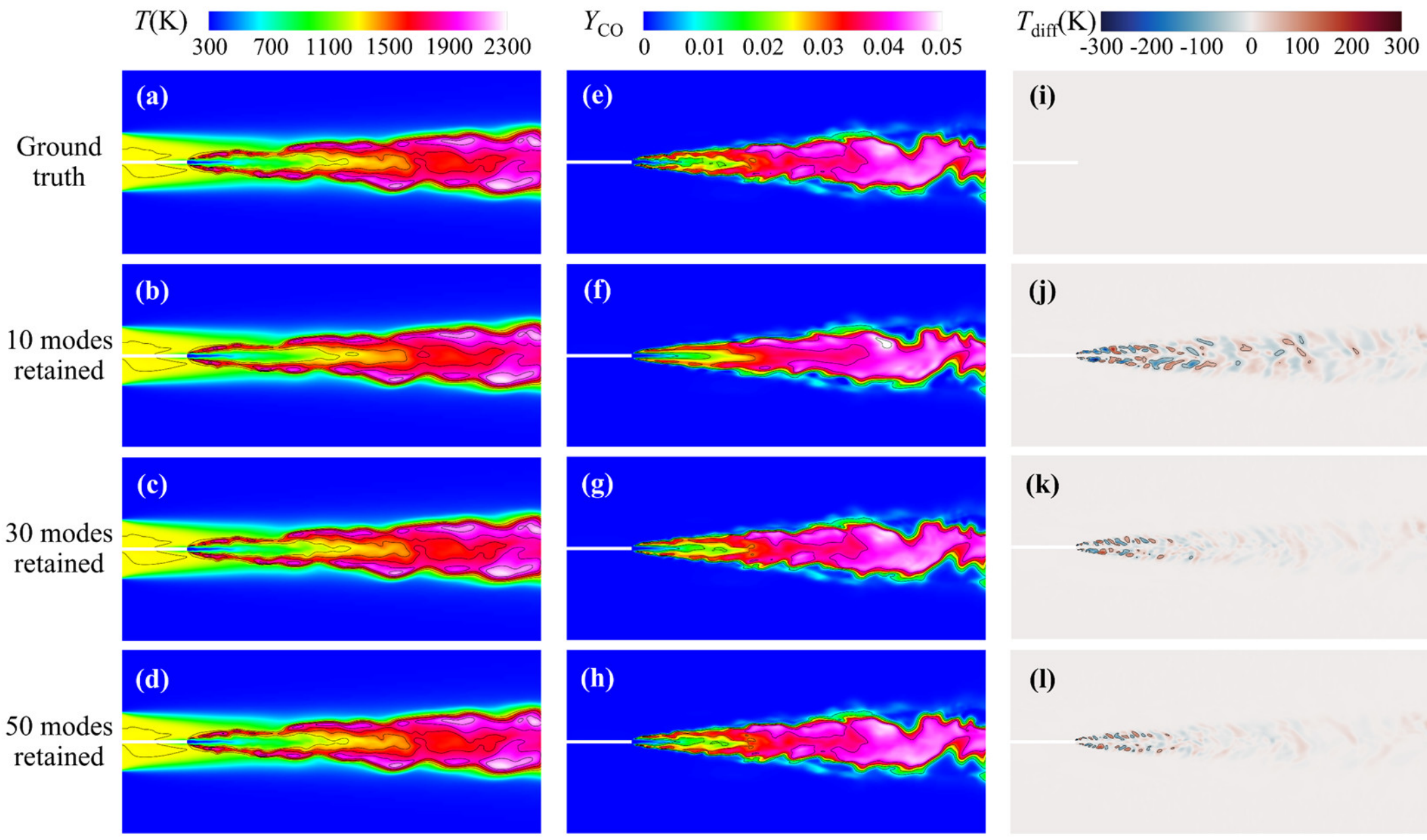

**Fig. 9.** Comparison of flow field features between CFD results and the POD reconstructions retaining 10, 30 and 50 most energetic modes for (a-d) temperature, (e-h) CO mass fraction, and (i-l) differences of reconstructed temperature field.

Additionally, the RRMSE between the reconstructions and original LES results is shown in Fig. 10. As the number of retained modes increases, the RRMSE decreases for both temperature and the CO mass fraction, indicating improved reconstruction accuracy. Notably, the RRMSE of the CO mass fraction is higher than that of the temperature field, which can be attributed to the finer spatial structures and higher variability present in CO distributions.

In the present model, the reconstructed fields were used as training data. As such, some prediction error is introduced, partly due to the exclusion of less energetic POD modes. However, this exclusion is deliberate. The objective of the ROM is not to replicate every detail of the turbulent field, but rather to capture the dominant energy-containing dynamics that govern the essential behavior of the system. These dynamics are typically associated with coherent structures and large-scale flow features that play a critical role in combustion processes. Conversely, the less energetic modes often correspond to small-scale turbulence, stochastic fluctuations, or even numerical noise. Attempting to learn these features through a neural network would not only increase computational cost but also risk destabilizing the auto-regressive framework, leading to premature divergence in long-term predictions. By focusing on dominant modes, the present model emphasizes physical robustness and generalizability, which are crucial for predictive modeling in reactive flows and other complex multi-physics applications. This strategic trade-off ensures that the ROM remains computationally efficient while still delivering meaningful insights for engineering analysis and design.

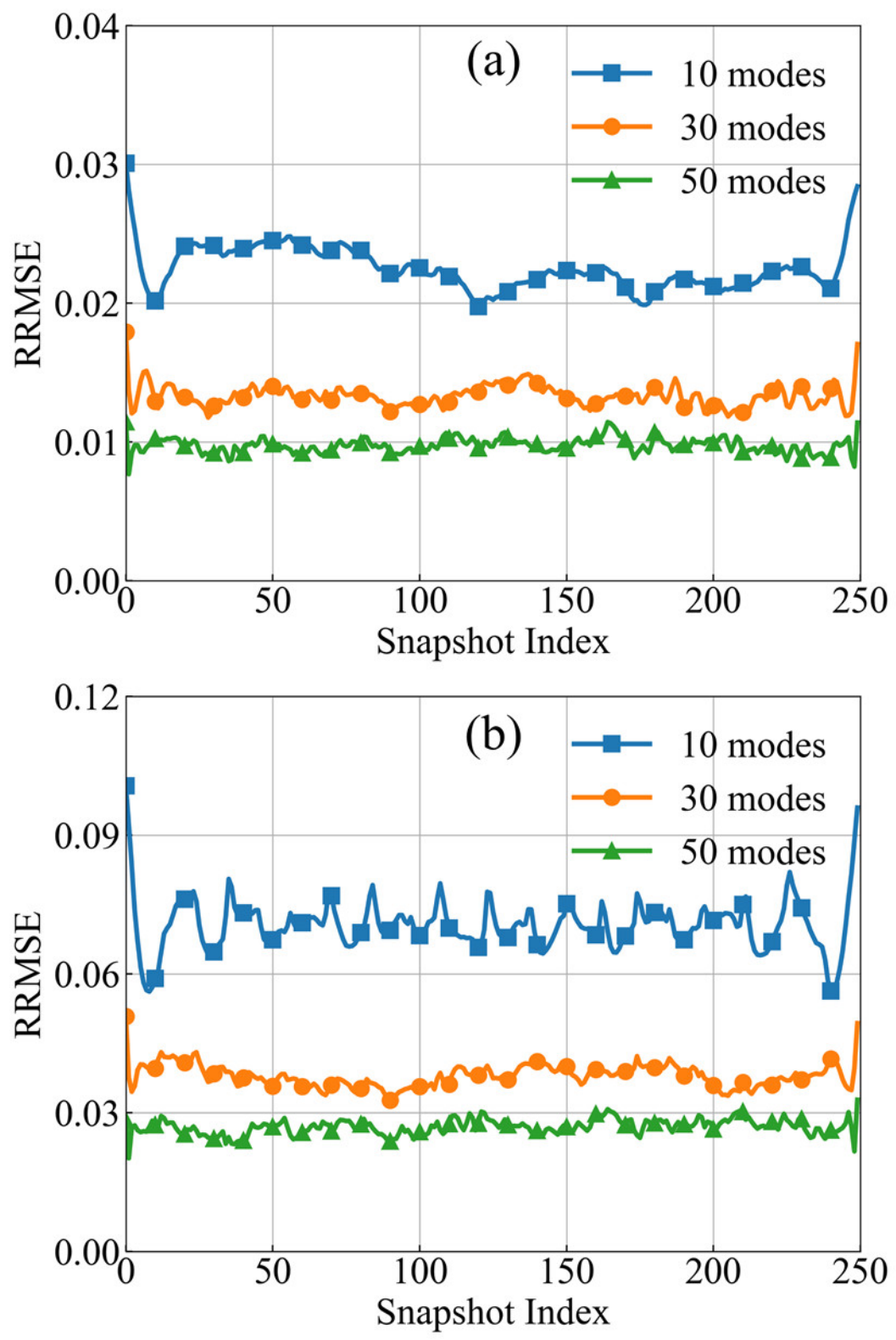

**Fig. 10.** RRMSE between LES results and reconstructions with 10, 30, 50 modes for representative variables: (a) temperature and (b) mass fraction of CO.

*4.3. Hybrid generative model to accelerate CFD*

The predictive capability of the POD-DL model for transient combustion field is evaluated in this section. Three cases were considered, with their respective parameters listed in Table 1. The prediction of such combustion fields presents significantly challenges due to the need to capture the evolution of small-scale structures and fluctuations. Firstly, the predictions were performed based on snapshots with time interval of $2\times10^{-4}$ s (Cases 01 and 02). A total number of 500 snapshots were extracted from LES results. The first 250 snapshots were used for training, while the remaining 250 snapshots were utilized as testing dataset. Temporal modes were obtained via POD by retrieving either 30 or 50 modes from the training dataset. These modes served as inputs to the deep learning model. Figure 11 compares the predicted results (lower part in each subfigure) with the original LES data (upper part in each subfigure) for temperature and the mass fractions of CO, $CO_2$ and $H_2O$. The predictions with 50 retained modes successfully capture the dominant spatial structures. However, there exists slightly difference between the isoline contours, which can be attributed to the chaotic properties of the current combustion filed. Prediction accuracy is higher in the upstream region of the jet, where the flow is more organized. Based on FFT analysis, the dominant frequencies of flow fluctuations decrease along the axial direction, indicating it is needed for finer temporal resolution or more training data snapshots to improve downstream accuracy. However, due to the computational cost of LES and the large size of individual snapshot, further data generation was not pursued in this work.

**Table 1** Summary of prediction cases with different time intervals and retained modes.

| Case | Time interval (s) | Retained modes | Training dataset | | Testing dataset | |
|---|---|---|---|---|---|---|
| | | | Time (s) | Snapshots | Time (s) | Snapshots |
| 01 | $2\times10^{-4}$ | 30 | (0.4, 0.45] | (0, 250] | (0.45, 0.5] | (250, 500] |
| 02 | $2\times10^{-4}$ | 50 | (0.4, 0.45] | (0, 250] | (0.45, 0.5] | (250, 500] |
| 03 | $1\times10^{-3}$ | 30 | (0.36, 0.46] | (0, 100] | (0.46, 0.56] | (100, 200] |

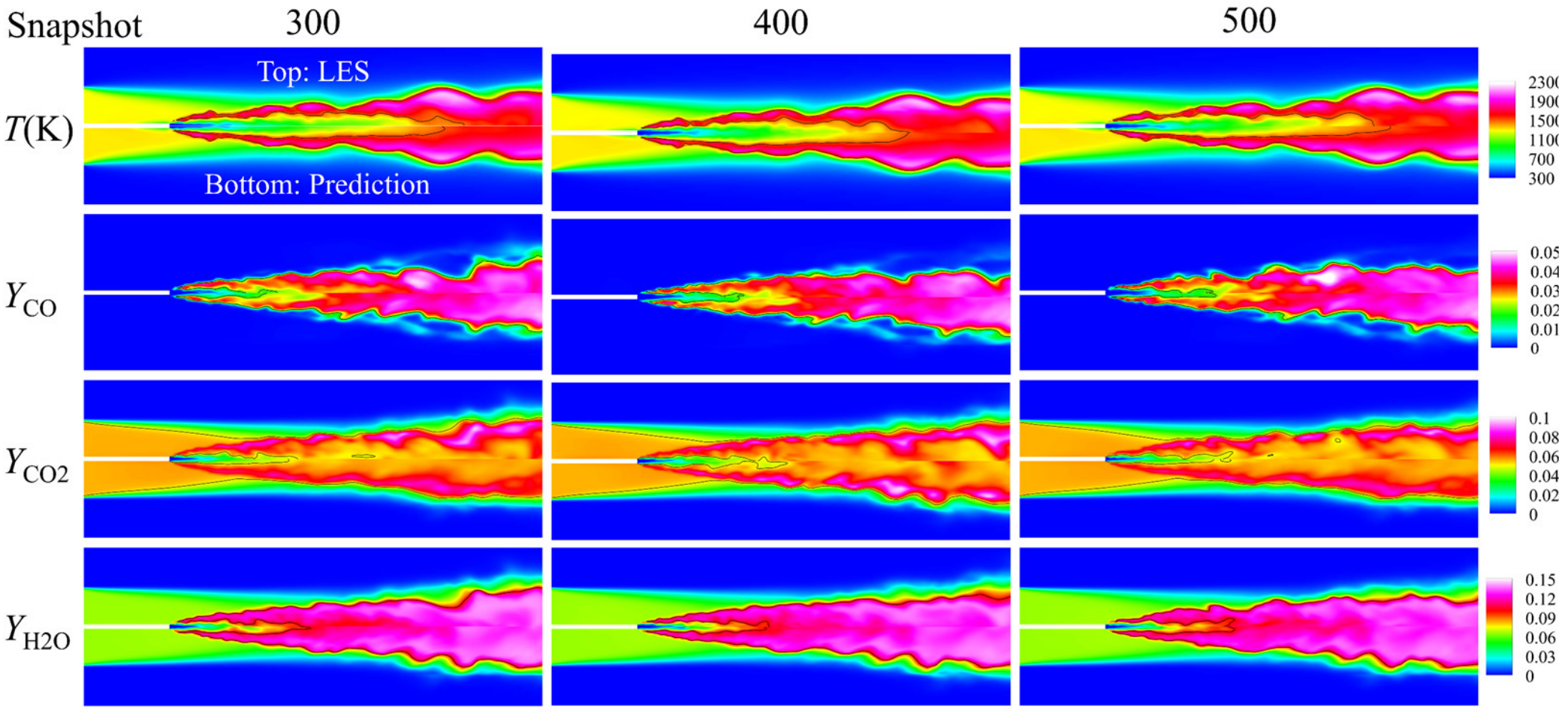

**Fig. 11.** Comparison of instantaneous fields from LES (top part in each subfigure) and hybrid ROM prediction with 50 modes (bottom part in each subfigure) for temperature, mass fractions of CO, $CO_2$ and $H_2O$ at several representative time instants.

The predicted radial distributions of temperature and species mass fractions with 30 and 50 modes (Cases 01 and 02) retained at axial positions $Z/D$ = 23.5 and 47 are compared with LES results, as presented in Fig. 12. Good agreement is achieved between predictions and original CFD results for temperature and $H_2O$ mass fraction. Regarding mass fractions of CO and $CO_2$, the POD-DL model slightly underpredicts or overpredicts the distributions at some time instants. As can be seen from the contours in Fig. 11, more fine structures exist in the distributions of CO and $CO_2$, resulting in difficulty in predictions. Notably, increasing the number of retained modes from 30 to 50 does not lead to significant improvements in predictive accuracy for these variables.

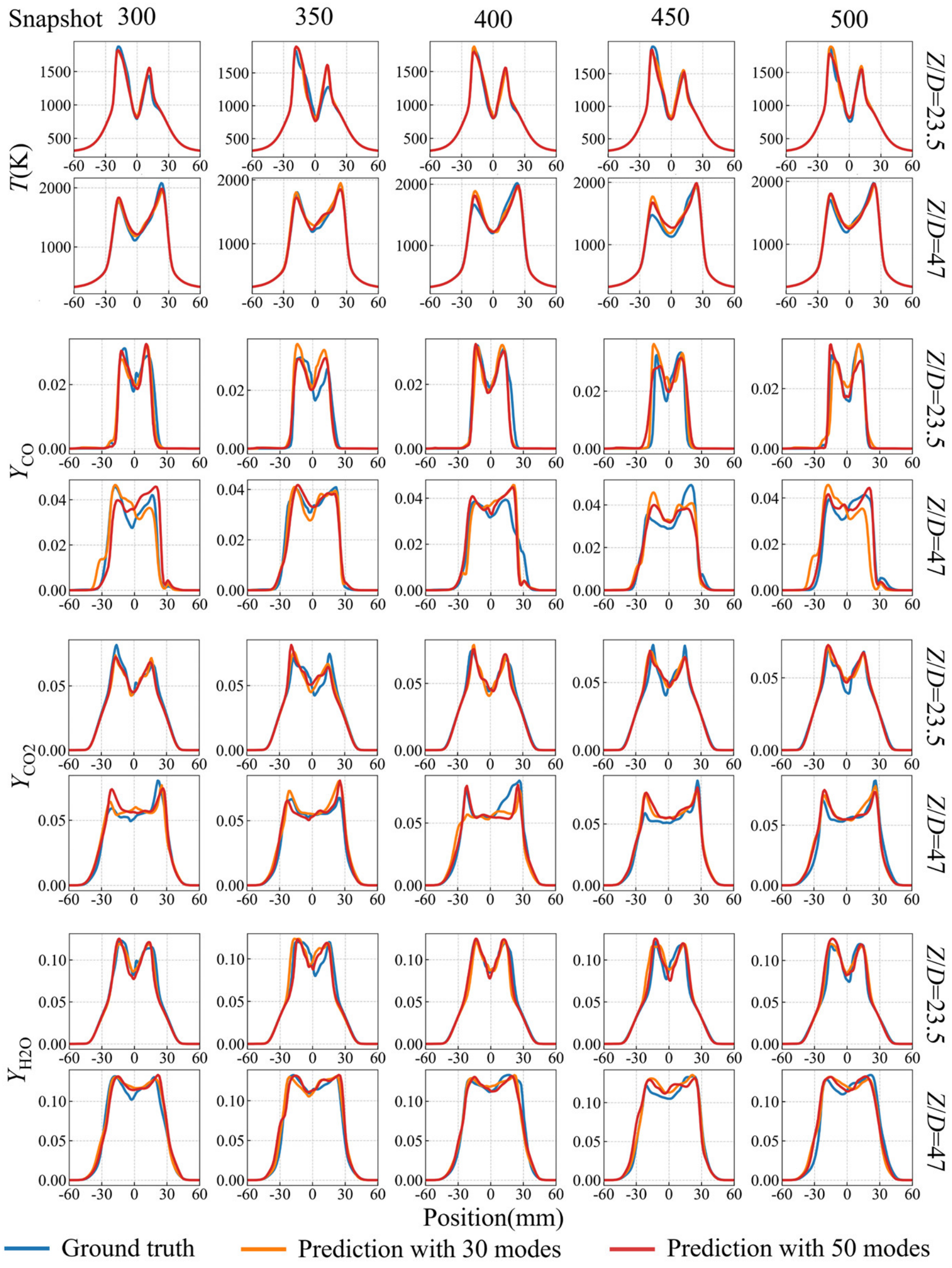

**Fig. 12.** Comparison of predicted and LES radial profiles of temperature and mass fractions of CO, $CO_2$,

and $H_2O$ at $Z/D$ = 23.5 and 47 at representative time instants using snapshot interval of $2\times10^{-4}$ s.

To quantitatively characterize the prediction accuracy, RRMSE defined in Eq. (26) was calculated for representative variables as a function of snapshot index, as shown in Fig. 13. And the RRMSE for specific snapshot is listed in Table 2. The lowest RRMSE is observed for temperature, showing a value around 0.05, whereas it increases for the predictions of $H_2O$, $CO_2$, and CO mass fractions. The discrepancy increases since the fine structures dominate as can be seen in the contours in Fig. 11.

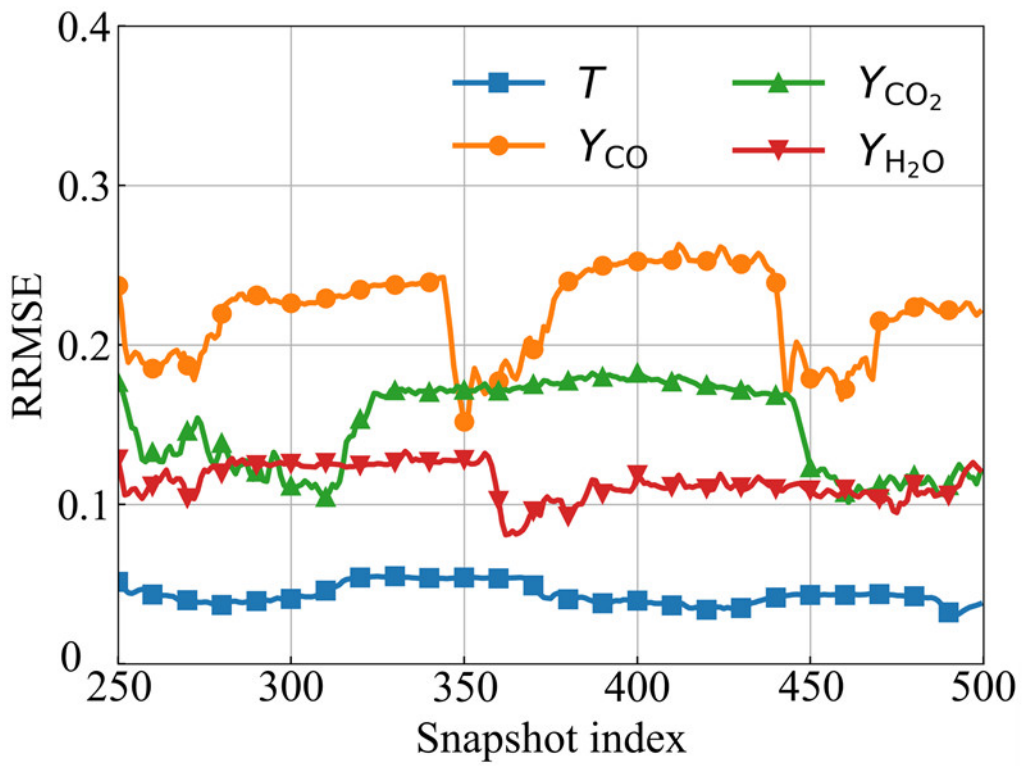

**Fig. 13.** RRMSE between predicted and LES results varying with snapshots for temperature, mass fractions of CO, $CO_2$ and $H_2O$.

**Table 2** RRMSE between predicted and LES results for various variables.

| Variables | Snapshot | | | | |
|---|---|---|---|---|---|
| | 300 | 350 | 400 | 450 | 500 |
| $T$ | 0.0465 | 0.0614 | 0.0435 | 0.0578 | 0.0451 |
| $Y_{CO}$ | 0.2652 | 0.1920 | 0.2593 | 0.2093 | 0.2454 |
| $Y_{CO2}$ | 0.1454 | 0.1865 | 0.1799 | 0.1529 | 0.1471 |
| $Y_{H2O}$ | 0.1320 | 0.1225 | 0.1351 | 0.1324 | 0.1328 |

Additionally, predictions were performed using a larger snapshot interval of $1\times10^{-3}$ s (Case 03) to examine the influence of reduced temporal resolution. Figure 14 compares radial distributions of temperature and mass fractions of CO, $CO_2$ and $H_2O$ at $Z/D$ = 23.5 and 47 for selected time instants. Temperature predictions remain accurate up to snapshot of 160, but slight deviations appeared thereafter, with overprediction at $Z/D$ = 23.5 and underprediction at $Z/D$ = 47. For species mass fractions, the predicted trends generally match the LES results, although differences in peak magnitudes and profile width remained.

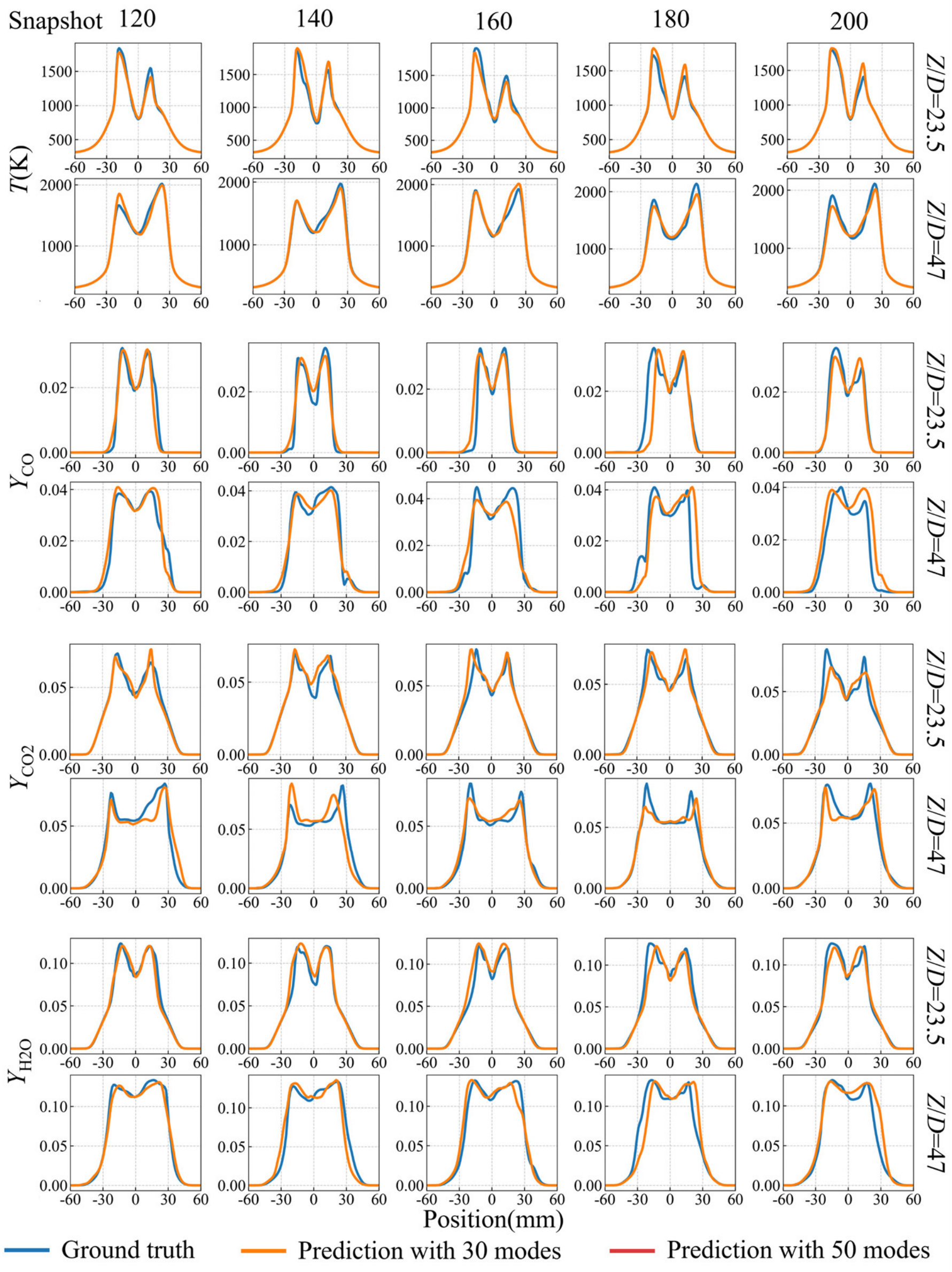

**Fig. 14.** Comparison of radial distributions of temperature and mass fractions of CO, $CO_2$, and $H_2O$ at axial positions *Z*/*D* = 23.5 and 47 based on snapshot interval of $1\times10^{-3}$ s.

It is worth noting that increasing the snapshot interval naturally acts as a temporal filter, suppressing small-scale fluctuations and emphasizing large-scale coherent structures. This filtering effect simplifies the prediction task by reducing the influence of high-frequency, small-scale turbulence that is often difficult to capture. Consequently, the model becomes more robust in forecasting the dominant flow dynamics over longer time horizons. A similar strategy is adopted in state-of-the-art generative forecasting models such as Google's GenCast [44], which uses coarser time intervals (e.g., 12 hours) to predict weather evolution over extended periods (e.g., 15 days). In this study, fewer data samples are available to train the model, which may compromise prediction accuracy to some extent. Nonetheless, the approach improves generalizability and computational efficiency for complex, multi-physics scenarios such as combustion.

RRMSE of Case 03 is shown in Fig. 15 and Table 3 lists the RRMSE for specific snapshot. Specifically, the RRMSE of CO mass fraction field that owns more fine structures slightly decreases as the snapshot interval increases. This indicates that reducing the temporal resolution omits the small-scale, and slightly decreases the RRMSE. Overall, prediction accuracy for the increased snapshot interval is comparable to that with the smaller snapshot interval, demonstrating the robustness of the present model with respect to temporal resolution.

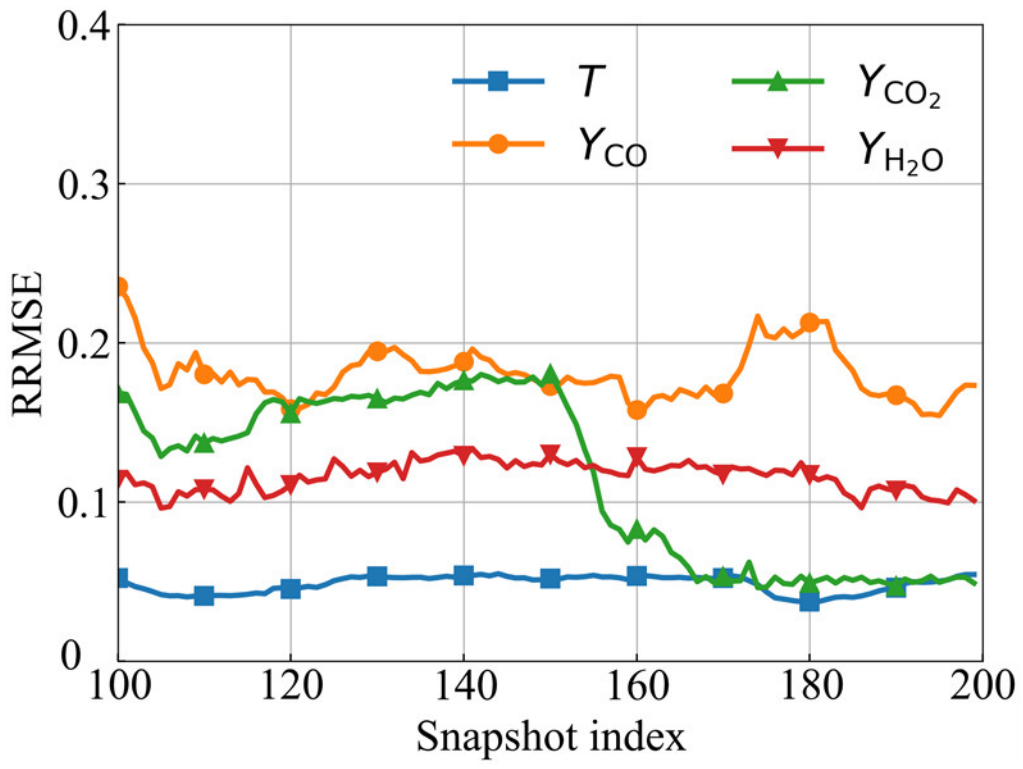

**Fig. 15.** RRMSE between predicted and LES results varying with snapshots for temperature, mass fractions of CO, $CO_2$ and $H_2O$.

**Table 3** RRMSE between predicted and LES results for various variables.

| Variables | Snapshot | | | | |
|---|---|---|---|---|---|
| | 120 | 140 | 160 | 180 | 200 |
| $T$ | 0.0503 | 0.0521 | 0.0507 | 0.0462 | 0.0560 |
| $Y_{CO}$ | 0.1941 | 0.1933 | 0.1724 | 0.2095 | 0.1800 |
| $Y_{CO2}$ | 0.1679 | 0.1671 | 0.1420 | 0.1617 | 0.1602 |
| $Y_{H2O}$ | 0.1156 | 0.1318 | 0.1182 | 0.1219 | 0.1077 |

To further assess the model performance, histograms of temperature and species mass fractions were compared between the predictions and original LES data, as shown in Fig. 16. Regarding the temperature, good match is observed for most temperature range, with slightly underrepresented counts near 1500 K and overrepresented values near 1700 K. For species mass fractions, the trend of distribution is well predicted though some difference is observed. This may be caused by the POD reconstruction and prediction.

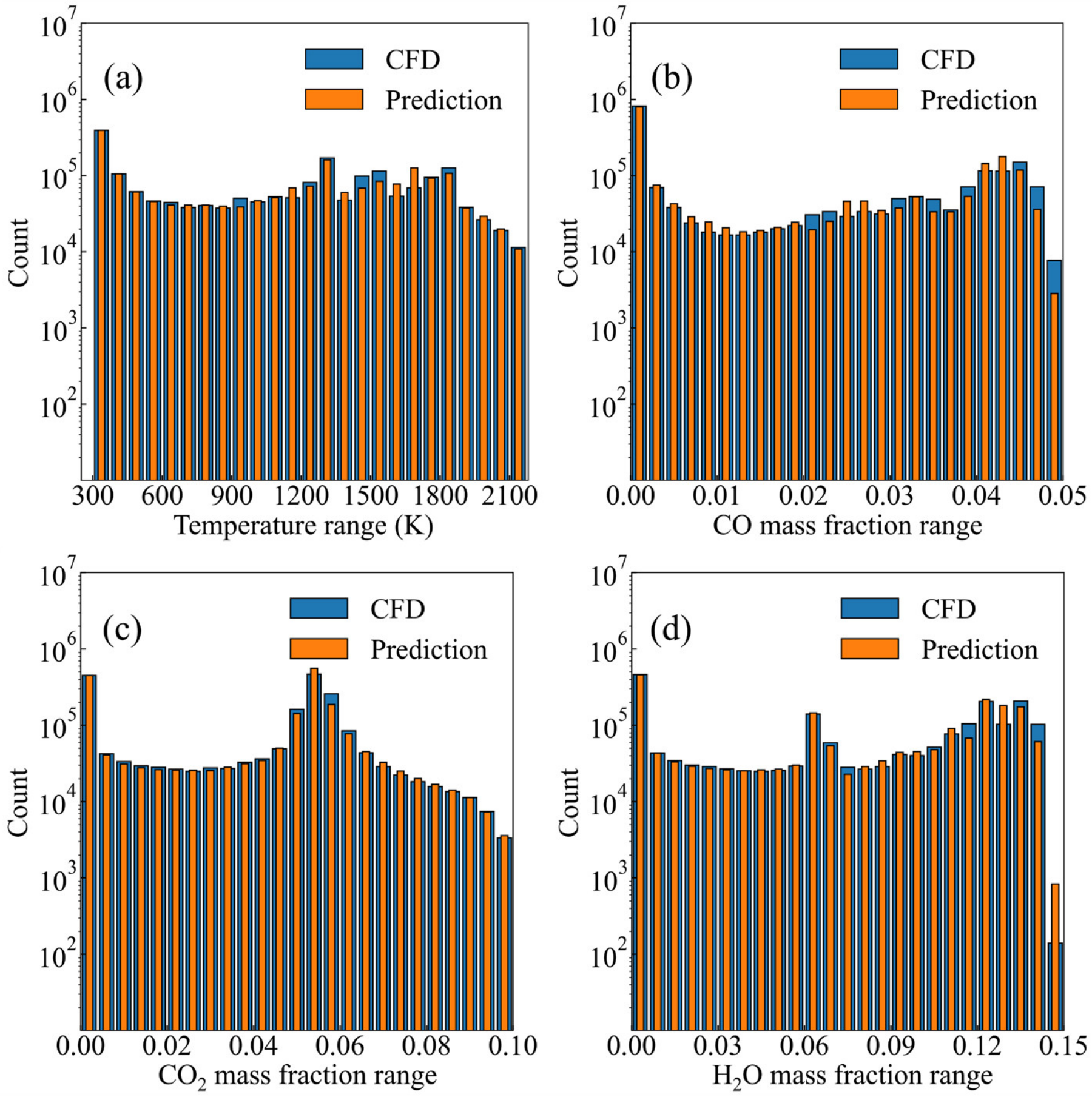

**Fig. 16.** Histograms of temperature, mass fractions of CO, $CO_2$ and $H_2O$ comparing between the predictions and LES results.

Finally, a computational cost comparison was conducted between LES and POD-DL model prediction. Both computations were performed on the same workstation equipped with 64 CPU cores (Dual Intel Xeon Platinum 8378A CPUs @ 3.00 GHz, 32 cores per CPU) and 256 GB Ram. LES used 60 CPU cores, while the prediction of POD-DL model was executed in parallel with one core per variable. Thus, all the variables can be predicted simultaneously. In terms of comparison, the physical flow time was kept the same for prediction and LES, namely 0.05 s and 0.1 s for Cases 01 and 03, respectively.

Figure 17 (a) plots the computational time of Cases 01 and 03 required for training/predicting for various representative variables. It is observed that the time cost for training and predicting are relatively uniform across species, but higher for temperature and velocity. The maximum training/predicting time (i.e., velocity in this study) was employed to compare with the LES computation time. For Cases 01 and 03, the training/predicting times of velocity component are respectively 1500.45 s and 430.12 s, the corresponding computational times of LES are respectively 181800 s and 363600 s, as shown in Fig. 17 (b). The computational time for LES is significantly larger than that of training/predicting time of POD-DL model. The main reason is that the time step for LES is $1\times10^{-6}$ s to keep a reasonable Courant number. In contrast, predictions based on POD-DL model operate at much larger time intervals. To further examine the efficiency of speed-up, the speed-up ratio (SUR) is defined, which is the ratio of computational times of LES and POD-DL model for predicting the same physical flow time. In this work, the SURs are approximately 121 and 845 for Cases 01 and 03, respectively. These results demonstrate that a significant computational acceleration is achieved using the present POD-DL model.

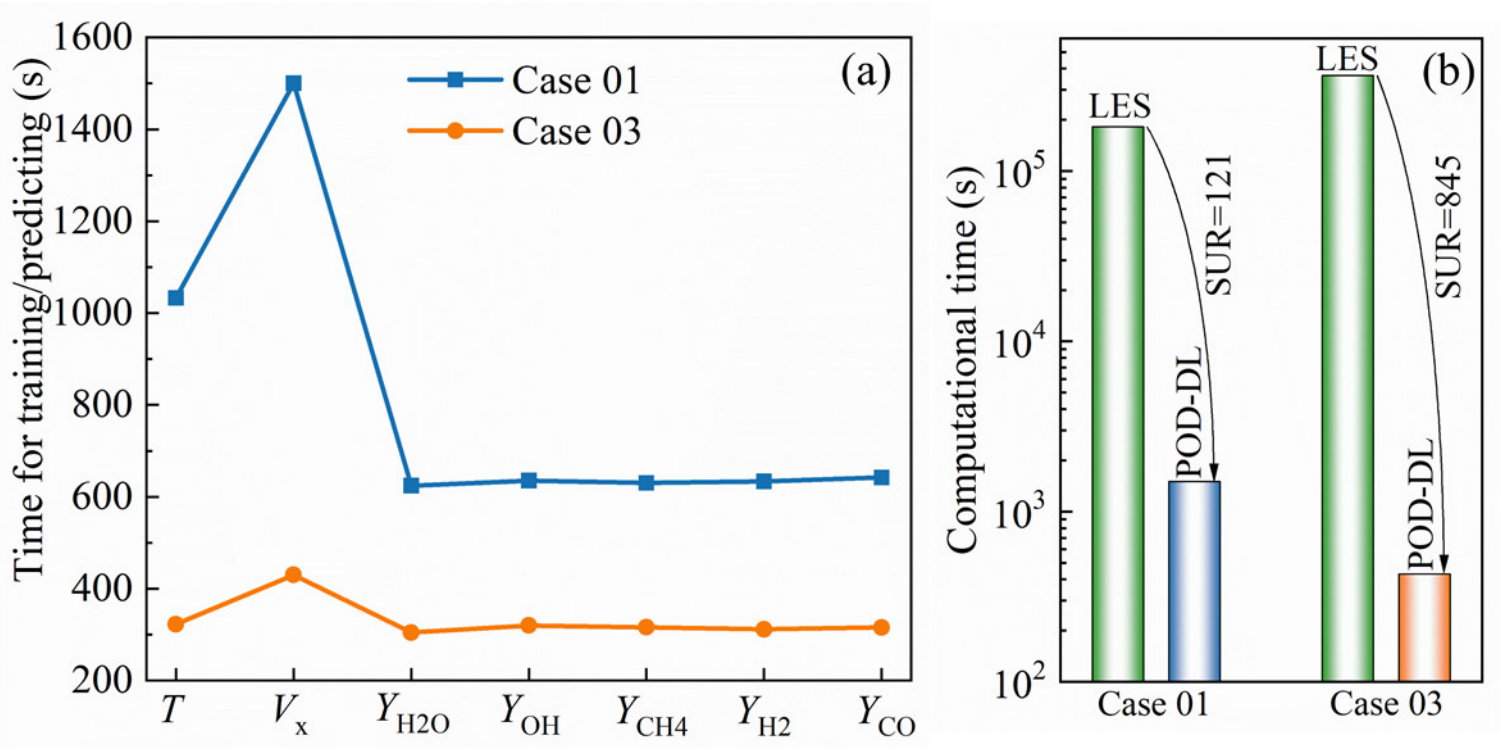

**Fig. 17.** (a) Computational time for training/predicting of several variables; (b) Computational time comparison of LES and POD-DL model for Cases 01 and 03, as well as the corresponding speed-up ratio.

## 5. Conclusions

This study aims to predict turbulent combustion dynamics with generative machine learning. A representative flame configuration, namely a jet-in-hot-coflow burner, was selected as a benchmark. Large eddy simulation (LES) was performed using the eddy dissipation concept model in combination with a skeletal mechanism comprising 17 species and 58 reactions. Reasonable agreement between simulations and experimental data fully validates the method. The LES results reveal that hydrogen/methane fuels diffuse radially outward and ignite in the shear mixing layer, forming a slightly lifted flame with an inverted cone shape.

A hybrid auto-regressive model, combining proper orthogonal decomposition with deep learning (POD-DL), was established to predict the temporal evolution of temperature and species fields. The model was trained on modal coefficients extracted from LES and evaluated on multiple test cases. Comparisons of instantaneous contours, radial profiles, histograms, and relative root mean square error demonstrated a reasonable agreement between the predictions and LES data, confirming the ability of the model to replicate the main flow features. Crucially, the hybrid POD-DL model significantly reduced computational cost, achieving speed-up ratios of 121 and 845 for the two tested temporal resolutions. This demonstrates the practical potential of generative AI to accelerate the CFD, particularly in reactive flow simulations where full LES remains computationally expensive. The framework provides a promising foundation for future efforts in physics-informed, data-driven modeling of multi-physics problems in combustion science.

## Acknowledgment

The authors acknowledge the ENCODING project that has received funding from the European Union's Horizon Europe research and innovation programme under the Marie Sklodowska-Curie grant agreement No. 101072779. S.L.C. acknowledges the MODELAIR project that has received funding from the European Union's Horizon Europe research and innovation programme under the Marie Sklodowska-Curie grant agreement No. 101072559. The results of this publication reflect only the author's view and do not necessarily reflect those of the European Union. The European Union cannot be held responsible for them. The authors acknowledge the grant PLEC2022-009235 funded by MCIN/AEI/ 10.13039/501100011033 and by the European Union "NextGenerationEU"/PRTR and the grant PID2023-147790OB-I00 funded by MCIU/AEI/10.13039 /501100011033 /FEDER, UE.

## Appendix A

Kinetic mechanism for methane and hydrogen combustion with air [36].

**Table A1** Gas-phase mechanism used [36].

| Number | Reaction | *A* | *b* | *Ea* |
|---|---|---|---|---|
| 1 | CH4(+M)=CH3+H(+M)<br>/0.0063 0.0 18000.0/<br>TBE H2O=5.0 | 6.3E14 | 0.0 | 104000 |
| 2 | CH4+O2=CH3+HO2 | 7.9E13 | 0.0 | 56000 |
| 3 | CH4+H=CH3+H2 | 2.2E4 | 3.0 | 8750 |
| 4 | CH4+O=CH3+OH | 1.6E6 | 2.36 | 7400 |
| 5 | CH4+OH=CH3+H2O | 1.6E6 | 2.1 | 2460 |
| 6 | CH3+O=CH2O+H | 6.8E13 | 0.0 | 0 |
| 7 | CH3+OH=CH2O+H2 | 1.0E12 | 0.0 | 0 |
| 8 | CH3+OH=CH2+H2O | 1.5E13 | 0.0 | 5000 |
| 9 | CH3+H=CH2+H2 | 9.0E13 | 0.0 | 15100 |
| 10 | CH2+H=CH+H2 | 1.4E19 | -2.0 | 0 |
| 11 | CH2+OH=CH2O+H | 2.5E13 | 0.0 | 0 |
| 12 | CH2+OH=CH+H2O | 4.5E13 | 0.0 | 3000 |
| 13 | CH+O2=HCO+O | 3.3E13 | 0.0 | 0 |
| 14 | CH+O=CO+H | 5.7E13 | 0.0 | 0 |
| 15 | CH+OH=HCO+H | 3.0E13 | 0.0 | 0 |
| 16 | CH+CO2=HCO+CO | 3.4E12 | 0.0 | 690 |
| 17 | CH2+CO2=CH2O+CO | 1.1E11 | 0.0 | 1000 |
| 18 | CH2+O=CO+H+H | 3.0E13 | 0.0 | 0 |
| 19 | CH2+O=CO+H2 | 5.0E13 | 0.0 | 0 |
| 20 | CH2+O2=CO2+H+H | 1.6E12 | 0.0 | 1000 |
| 21 | CH2+O2=CH2O+O | 5.0E13 | 0.0 | 9000 |
| 22 | CH2+O2=CO2+H2 | 6.9E11 | 0.0 | 500 |
| 23 | CH2+O2=CO+H2O | 1.9E10 | 0.0 | -1000 |
| 24 | CH2+O2=CO+OH+H | 8.6E10 | 0.0 | -500 |
| 25 | CH2+O2=HCO+OH | 4.3E10 | 0.0 | -500 |
| 26 | CH2O+OH=HCO+H2O | 3.43E9 | 1.18 | -447 |
| 27 | CH2O+H=HCO+H2 | 2.19E8 | 1.77 | 3000 |
| 28 | CH2O+M=HCO+H+M | 3.31E16 | 0.0 | 81000 |
| 29 | CH2O+O=HCO+OH | 1.81E13 | 0.0 | 3082 |
| 30 | HCO+OH=CO+H2O | 5.0E12 | 0.0 | 0 |
| 31 | HCO+M=H+CO+M | 1.60E14 | 0.0 | 14700 |
| 32 | HCO+H=CO+H2 | 4.00E13 | 0.0 | 0 |
| 33 | HCO+O=CO2+H | 1.0E13 | 0.0 | 0 |
| 34 | HCO+O2=HO2+CO | 3.3E13 | -0.4 | 0 |
| 35 | CO+O+M=CO2+M | 3.20E13 | 0.0 | -4200 |
| 36 | CO+OH=CO2+H | 1.51E7 | 1.3 | -758 |
| 37 | CO+O2=CO2+O | 1.6E13 | 0.0 | 41000 |
| 38 | HO2+CO=CO2+OH | 5.80E13 | 0.0 | 22934 |
| 39 | H2+O2=2OH | 1.7E13 | 0.0 | 47780 |
| 40 | OH+H2=H2O+H | 1.17E9 | 1.3 | 3626 |
| 41 | H+O2=OH+O | 5.13E16 | -0.816 | 16507 |
| 42 | O+H2=OH+H | 1.8E10 | 1.0 | 8826 |
| 43 | H+O2+M=HO2+M | 3.61E17 | -0.72 | 0 |
| | TBE H2O=18.6 CO2=4.2 | | | |
| | H2=2.86 CO=2.11 N2=1.26 | | | |
| 44 | OH+HO2=H2O+O2 | 7.5E12 | 0.0 | 0 |
| 45 | H+HO2=2OH | 1.4E14 | 0.0 | 1073 |
| 46 | O+HO2=O2+OH | 1.4E13 | 0.0 | 1073 |
| 47 | 2OH=O+H2O | 6.0E8 | 1.3 | 0 |
| 48 | H+H+M=H2+M | 1.0E18 | -1.0 | 0 |
| 49 | H+H+H2=H2+H2 | 9.2E16 | -0.6 | 0 |
| 50 | H+H+H2O=H2+H2O | 6.0E19 | -1.25 | 0 |
| 51 | H+H+CO2=H2+CO2 | 5.49E20 | -2.0 | 0 |
| 52 | H+OH+M=H2O+M | 1.6E22 | -2.0 | 0 |
| | TBE H2O=5.0 | | | |
| 53 | H+O+M=OH+M | 6.2E16 | -0.6 | 0 |
| | TBE H2O=5.0 | | | |
| 54 | H+HO2=H2+O2 | 1.25E13 | 0.0 | 0 |
| 55 | HO2+HO2=H2O2+O2 | 2.0E12 | 0.0 | 0 |
| 56 | H2O2+M=OH+OH+M | 1.3E17 | 0.0 | 45500 |
| 57 | H2O2+H=HO2+H2 | 1.6E12 | 0.0 | 3800 |
| 58 | H2O2+OH=H2O+HO2 | 1.0E13 | 0.0 | 1800 |

Note: Reaction rate constants are in form $k_{\mathrm{f}} = AT^{\mathrm{b}} \exp\left(-Ea/RT\right)$. Units are moles, cm, s, K and cal/mol.